\documentclass[prd,showpacs,amsmath,%showkeys,
twocolumn,floatfix,amssymb, preprintnumbers, nofootinbib, superscriptaddress]{revtex4} 
\usepackage{hyperref}
\usepackage{epsfig,dcolumn}
\usepackage{graphicx}
\usepackage{comment} 
\usepackage{verbatim}
\usepackage{color}
\usepackage{graphicx}
\usepackage{bm}
\usepackage{ifpdf}
\usepackage{multirow}
\usepackage{soul} % \st{} - cross the phrase

\def\lsim{\mathrel{\rlap{\lower4pt\hbox{\hskip1pt$\sim$}}
    \raise1pt\hbox{$<$}}}
\def\gsim{\mathrel{\rlap{\lower4pt\hbox{\hskip1pt$\sim$}}
    \raise1pt\hbox{$>$}}}
\begin{document}

\title{Three-body Final State Interaction in \mbox{$\eta \to 3 \pi$}}

\author{Peng~Guo}
\email{pguo@jlab.org}
\affiliation{Center for Exploration of Energy and Matter, Indiana University, Bloomington, IN 47403}
\affiliation{Physics Department, Indiana University, Bloomington, IN 47405, USA}
\affiliation{Thomas Jefferson National Accelerator Facility, %12000 Jefferson Avenue,  
Newport News, VA 23606, USA}
\author{Igor~V.~Danilkin}
\affiliation{Thomas Jefferson National Accelerator Facility, %12000 Jefferson Avenue,  
Newport News, VA 23606, USA}
\author{Diane~Schott}
\affiliation{Thomas Jefferson National Accelerator Facility, %12000 Jefferson Avenue,  
Newport News, VA 23606, USA}
\affiliation{Department of Physics, The George Washington University, Washington, DC 20052, USA}
\author{C.~Fern\'andez-Ram\'{\i}rez}
\affiliation{Thomas Jefferson National Accelerator Facility, %12000 Jefferson Avenue,  
Newport News, VA 23606, USA}
\author{V.~Mathieu}
\affiliation{Center for Exploration of Energy and Matter, Indiana University, Bloomington, IN 47403}
\affiliation{Physics Department, Indiana University, Bloomington, IN 47405, USA}
\author{Adam~P.~Szczepaniak}
\affiliation{Center for Exploration of Energy and Matter, Indiana University, Bloomington, IN 47403}
\affiliation{Physics Department, Indiana University, Bloomington, IN 47405, USA}
\affiliation{Thomas Jefferson National Accelerator Facility, %12000 Jefferson Avenue,  
Newport News, VA 23606, USA}
 
\collaboration{Joint Physics Analysis Center}
\preprint{JLAB-THY-15-2041}

\date{\today}

\begin{abstract} 
We present a unitary dispersive model for the $\eta \to 3 \pi$ decay process based upon the Khuri-Treiman equations which are solved by means of the Pasquier inversion method. The description of the hadronic final-state interactions for the $\eta \to 3\pi$ decay is essential to reproduce the available data and to understand the existing discrepancies between Dalitz plot parameters from experiment and chiral perturbation theory. Our approach incorporates substraction constants that are fixed by fitting the recent high-statistics WASA-at-COSY data for $\eta \to \pi^+ \pi^- \pi^0$. Based on the parameters obtained we predict the slope parameter for the neutral channel to be $\alpha=-0.022\pm 0.004$. Through matching to next-to-leading-order chiral perturbation theory we estimate the quark mass double ratio to be $Q=21.4 \pm 0.4$. 
\end{abstract} 


% 13.25.Jx - Decays of other mesons (section: Hadronic decays of mesons)
% 11.55.Fv - Dispersion relations (section: S-matrix-theory; analytic structure of amplitudes)
% 14.65.Bt - Light quarks (section: properties of specific particles)
% 12.39.Fe - Chiral Lagrangians (section: Phenomenological quark models)
%http://www.aip.org/publishing/pacs/pacs-reg10#14

\maketitle

\section{Introduction}\label{intro}
Production of  three particles plays an important role in hadron physics. It sheds light on the reaction dynamics, {\it e.g.} the OZI rule, and can amplify production of hadron resonances, with the mysterious XYZ states seen in the spectrum of charmonia and  bottomonia \cite{PDG-2012} being the most recent examples.  The need for precision analysis of final states containing three light hadrons has become even more pressing given the high quality data emerging from the various hadron facilities around the world, including Jefferson Lab, COMPASS and BESIII \cite{Battaglieri:2010zza,Eugenio:2003,Adolph:2014rpp, Ablikim:2015cmz}. Recently, significant progress has been made in analysis of hadron-hadron interactions at low energies based on the S-matrix principles of unitarity, analyticity and crossing symmetry \cite{Ananthanarayan:2000ht, GarciaMartin:2011cn, Buettiker:2003pp, Gasparyan:2010xz}. At low energies, unitarity is an important constraint given that there is only a limited number of contributing channels. Unitarity also determines the analytical properties of partial waves and constraints resonant scattering. Implementation of crossing-symmetry is much more difficult since it is related to the underlying dynamics.  However, at low energies it can be systematically investigated by identifying the most important, {\it i.e.} closest to the physical region, singularities of the cross-channel  amplitudes, and for example in reactions involving Goldstone bosons these can be constrained by chiral symmetry of QCD~\cite{Weinberg:1967tq, Gasser:1983yg}.

In this paper we focus on decays of  the $\eta$ meson to three pions. From the experimental side, the high-quality data from WASA-at-COSY \cite{Adlarson:2014aks,Bashkanov:2007aa}, Crystal Barrel \cite{Abele:1998yj,Tippens:2001fm}, and KLOE \cite{Ambrosino:2008ht,Ambrosinod:2010mj}, along with the data from CLAS \cite{Eugenio:2003},  which is currently being analyzed, present an opportunity for precision analysis of the Dalitz distribution. In the charged decay channel, $\eta\to \pi^+\pi^-\pi^0$, we only have access to the binned data from the  WASA-at-COSY \cite{Adlarson:2014aks} experiment and therefore it is the only data set we use in our data-driven analysis. From the theoretical point of view $\eta\to 3\pi$ decays  are of interest because of isospin  violation. These decays are dominated by the  intrinsic isospin breaking effects in QCD as electromagnetic effects are expected to be small \cite{Sutherland:1966zz,Bell:1996mi}. Consequently, the decay width for \mbox{$\eta\rightarrow 3\pi$} is expected to be proportional to the light quark mass difference and the decay amplitude is often expressed in terms of the quantity, $1/Q^2$ defined by 
\begin{equation}\label{Eq:Q}
%R^{-1}&=&\frac{m_d-m_u}{m_s-\hat m}\,,\\
\frac{1}{Q^2}=\frac{m_d^2-m_u^2}{m_s^2-\hat m^2}. 
\end{equation}
Here \mbox{$\hat m = (m_u+m_d)/2$} is the average of the $u$ and $d$ quark masses. One determines $Q$ by comparing a theoretical prediction with the experimental decay width \mbox{$\Gamma(\eta\rightarrow\pi^+\pi^-\pi^0) = 281\pm28$ eV}  \cite{PDG-2012}. However, it  is important to emphasize that this procedure requires that the amplitude implements chiral constraints or at least it agrees with the leading-order chiral perturbation theory ($\chi$PT), which is where $Q$ originates. Once $Q$ is extracted, it can be combined with the knowledge of the  $\hat m$ and $m_s$, {\it e.g.} from lattice simulations, to determine the light-quark mass difference.

It is necessary to consider the $\eta \to 3\pi$ decay amplitudes beyond $\chi$PT. This is apparent when considering contributions to   \mbox{$\Gamma(\eta\rightarrow\pi^+\pi^-\pi^0)$} from  the first few terms in the low energy expansion. Specifically, the leading-order $\chi$PT result,  \mbox{$\Gamma^{\text{LO}}_{\eta\rightarrow \pi^+\pi^-\pi^0}=66$ eV} \cite{Cronin:1967jq,Osborn:1970nn}, is  approximately four times smaller than expected. Inclusion of next-to-leading (one loop) corrections increases the theoretical prediction to \mbox{$\Gamma^{\text{NLO}}_{\eta\rightarrow \pi^+\pi^-\pi^0}=167\pm50$ eV}  \cite{Gasser:1984pr}, which is still significantly below the data. The next-to-next-to-leading calculation (two loops) has been performed recently \cite{Bijnens:2007pr}. It pushes the decay width further towards the data; however, it contains a large number of low energy constants. In addition to the apparent poor convergence, low orders of $\chi$PT  give an incorrect result for the shape of the Dalitz distribution in the neutral $3\pi^0$ decay. To the leading order, this distribution is represented by a single parameter, $\alpha$, which $\chi$PT predicts to be positive while the experimental result  is  \mbox{$\alpha=-0.0317\pm0.0016$} \cite{PDG-2012}. The fact that chiral expansion converges slowly indicates the importance of final state interactions. This is expected to be a consequence of unitarity, which in $\chi$PT is incorporated only order by order. To fulfill unitarity various dispersive frameworks were developed  \cite{Kambor:1995yc, Anisovich:1996tx} with recent updates of \cite{Colangelo:2009db, Lanz:2013ku} and \cite{Descotes-Genon:2014tla}. These analyses are based on the Khuri-Treiman (KT) representation \cite{Khuri:1960kt}. In the KT approach, partial waves are given in the elastic approximation with the left-hand cut contributions computed from cross-channel amplitudes that are approximated by the same elastic partial waves as in the direct channel and are bootstrapped. Other calculations employed, for example, nonrelativistic effective field theory (NREFT) \cite{Schneider:2010hs} and alternative dispersive approaches were studied in \cite{Kampf:2011wr}.

The final state interactions in  $\eta  \to 3\pi$ at low energies can be approximated by elastic $\pi\pi$ scattering. These amplitudes are available with high precision up to \mbox{$\sqrt{s}=1.1$ GeV}  \cite{GarciaMartin:2011cn}. However, dispersion calculations involve integrals over all energies. In order to suppress the unknown high-energy region, the dispersive integrals are usually over-subtracted and the subtraction constants are fixed by comparing to the data \cite{Lanz:2013ku,Niecknig:2012sj}. In \cite{Danilkin:2014cra} the authors used an alternative method whereby the dispersive integral was split into elastic and inelastic contributions and the latter was parametrized by a power series in a suitably chosen conformal variable. In the current work, we apply yet a different approach. We obtain the solution of the  KT equation using the so-called Pasquier inversion method \cite{Pasquier:1968kt,Aitchison:1978pw}. In this case the dependence on the unknown high energy region is traded for by the dependence on the far left-hand cuts. The advantages and disadvantages of alternative procedures were discussed in \cite{Guo:2014vya}.

The paper is organized as follows. In Section \ref{SectionII} we present the basic formalism and discuss how three body effects are incorporated using the Pasquier inversion.  The numerical results are presented in Section \ref{SectionIII}, which we divide into two parts. In the first part we perform a data-driven dispersive analysis of the WASA-at-COSY data \cite{Adlarson:2014aks} without input from $\chi$PT. We show the fitted Dalitz plot parameters for the charged decay and predict the slope parameter for the neutral decay channel. In the second part we match our amplitudes to  $\chi$PT in order to extract the  $Q$ value. Conclusions are summarized in Section \ref{SectionIV}. 
  
\section{Formalism}
\label{SectionII}

\subsection{Kinematics and partial wave expansion}

The isospin violating $\eta \to 3\pi$ decay involves a $\Delta I =1$ interaction. The transition matrix elements, \mbox{$A^{\alpha \beta \gamma \eta}(s,t,u)$}, depends on four isospin indices, with the index $\eta$ referring to the isospin component of the interaction and $\alpha,\beta,\gamma$ to three pions.  In terms of the particle momenta the three Mandelstam variables are \mbox{$s=(p_1 + p_2)^2= (p_4-p_3)^2$}, \mbox{$t=(p_2 + p_3)^2 = (p_4-p_1)^2$}, and \mbox{$u = (p_1 + p_3)^2 = (p_4 - p_2)^2$}. The Mandelstam variables satisfy  \mbox{$s+t+u = m_\eta^2 + m_{1}^2+m_{2}^{2}+m_{3}^{2}$}, with $m_\eta$ being the mass of the $\eta$, also referred to as particle $i=4$  and $m_i$, $i=1.. 3$ to the pions. On account of crossing symmetry, the following processes are described by the same complex function (with the bar denoting an antiparticle): the $s$-channel scattering \mbox{$4 + \bar 3 \to 1 + 2 $}, the $t$-channel scattering \mbox{$4 + \bar 1 \to  2 + 3$}, the $u$-channel scattering, \mbox{$4 + \bar 2 \to 1 +  3$}, and the decay channel \mbox{$4\to  1 + 2  + 3$}.  In particular the amplitude in the decay channel will be derived by analytical continuation of the $s$-channel partial wave expansion.  In the $s$-channel, the  amplitude \mbox{$A^{ \alpha \beta \gamma \eta} (s,t,u)$} has the following partial wave (p.w.) decomposition, 
\begin{equation}\label{ExactPW}
A^{\alpha \beta \gamma \eta}(s,t,u) =\sum_{L=0}^{\infty}\sum_{I}(2L+1) P_{L}(z_{s})\,\mathcal{P}^{(I)}_{\alpha \beta \eta \gamma} A_{I L} (s)\,,
\end{equation}
where \mbox{$P_{L}(z_s)$} is the Legendre polynomial and $z_s$ is a cosine of the center-of-mass scattering angle $\theta_s$,
\begin{equation}\label{zs} 
z_s \equiv\cos\theta_s=\frac{s\,(t-u)+ (m_{1}^{2}-m_{2}^{2})\,(m_\eta^{2} -m_{3}^{2})}{ \lambda^{1/2}(s,m_\eta^{2} ,m_{3}^2)\,\lambda^{1/2}(s,m_{1}^2,m_{2}^2) }\,.
\end{equation} 
The usual K\"all\'en triangle function is given by \mbox{$\lambda(a,b,c) = a^2 + b^2 + c^2 - 2\,(a\,b + b\,c + a\,c)$} and \mbox{$(I,L)$} label isospin and orbital angular momentum quantum numbers in the $s$-channel with \mbox{$I+L=even$} due to Bose symmetry of pions. The isospin projection operators \mbox{$\mathcal{P}^{(I)}_{\alpha \beta \gamma \eta}$} are given in Appendix \ref{App:A}. We note that at this stage the partial waves are arbitrarily normalized. The unitary relation, which we discuss in the following is homogeneous in $A$  and at the end we will normalize the amplitude by comparing with the experimental data.

\begin{figure*}[t]
\includegraphics*[width=0.75\textwidth]{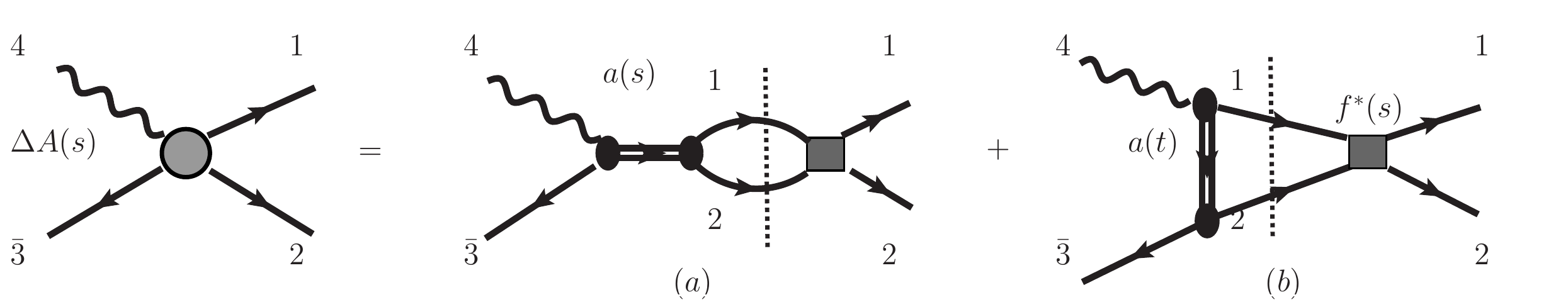}
\caption{A diagrammatic representation of discontinuity relations in Eq.\,(\ref{dis}). 
\label{fig:dis}}
\end{figure*}

The p.w. amplitudes $A_{I L} (s)$ have both the right-hand cut discontinuities demanded by the direct channel unitarity and left-hand cut discontinuities from exchanges in the $t$ and $u$ channels. We emphasize that Eq. (\ref{ExactPW}) is exact in the $s$-channel physical region, when the infinite sum over $L$ converges. The amplitudes in the other channels are obtained by analytical continuation. Low-energy approaches based on partial wave expansion involve truncation of the partial waves series at some $L=L_{max} < \infty$, which violates analytical properties of cross-channel amplitudes. To partially recover those, we represent the amplitude as a sum of truncated partial wave series in each of the three channels \cite{Khuri:1960kt,Bronzan:1963kt,Aitchison:1965kt,Aitchison:1965dt,Aitchison:1966kt,Pasquier:1968kt,Pasquier:1969dt},
\begin{align}\label{decayamp}
A^{\alpha \beta \gamma \eta}&(s,t,u) =\sum_{L=0}^{L_{max}}\sum_{I}(2L+1) \left(P_{L}(z_{s})\,\mathcal{P}^{(I)}_{\alpha \beta \eta \gamma} a_{I L} (s)\right. \nonumber \\
& \quad     \left. + P_{L}(z_{t})\,\mathcal{P}^{(I)}_{ \beta \gamma \alpha \eta }a_{I L} (t) + P_{L}(z_{u })\,     \mathcal{P}^{(I)}_{\gamma \alpha \beta \eta} a_{I L} (u) \right ),
\end{align}
where the amplitudes are $a_{IL}$ defined as having only right-hand discontinuities demanded by unitarity in the respective channels. The center of mass scattering angles in the $t$- and the $u$-channel are given by 
\begin{align} 
&  z_t  =  \frac{t\,(s - u)+ (m_{3}^{2}-m_{2}^{2})\,(m_\eta^{2} -m_{1}^{2})}{ \lambda^{1/2}(t,m_\eta^2,m_{1}^2)\,\lambda^{1/2}(t,m_{2}^2,m_{3}^2) }, \nonumber \\
&  z_u =  \frac{u\,(t - s)+ (m_{1}^{2}-m_{3}^{2})\,(m_\eta^{2} -m_{2}^{2}) }{\lambda^{1/2}(u,m_\eta^2,m_{2}^2)\, \lambda^{1/2}(u,m_{1}^2,m_{3}^2)}. 
\end{align}
We remark that the decomposition in Eq.\,(\ref{decayamp}) satisfies crossing symmetry explicitly; however,  violation of analyticity remains since the amplitude contains a finite number of  high-spin partial waves in any given channel. This would be a problem at high energies but hopefully does not influence our low-energy analysis. What the representation in Eq.~(\ref{decayamp}) does is to allow for unitarity to be implemented in all three channels. We also note that decomposition in  Eq.(\ref{decayamp})  is exact up to NNLO in $\chi$PT \cite{Stern:1993rg, Knecht:1995tr} and is often referred to as  ``reconstruction theorem".

It is convenient to express the p.w. amplitude \mbox{$A_{IL} (s)$} ({\it c.f.} Eq.~(\ref{ExactPW})) in terms of the amplitudes \mbox{$a_{IL}(s)$} that are defined by Eq.\,(\ref{decayamp}),
\begin{equation} \label{Eq:AIL}
A_{I L} (s)=a^{Right}_{IL}(s) + a^{Left}_{IL}(s). 
\end{equation} 
Here the amplitude $a^{Right}_{IL}(s)$ has only the right-hand discontinuity, 
\begin{equation} 
a^{Right}_{IL}(s)  = a_{IL}(s), 
\end{equation} 
and the left-hand discontinuities of $a^{Left}_{IL}(s)$ originate from the exchange terms, 
\begin{eqnarray}\label{Eq:AIL}
& & a^{Left}_{I L} (s) = \sum_{L'=0}^{L_{max}}\sum_{I'}\frac{(2L'+1)}{2} \int_{-1}^{1} 
dz_{s} P_{L}(z_{s})\nonumber\\
& & \times \big (P_{L'}(z_{t}) C_{st}^{I I'}  a_{I' L'}(t)+P_{L'}(z_{u }) 
C_{s u}^{I I'} a_{I' L'}(u) \big)\,. \nonumber \\
\end{eqnarray}
Here \mbox{$C_{st}$} and \mbox{$C_{su}$} are the standard crossing matrices and are given in Appendix \ref{App:A}.

\subsection{Unitarity and the three-body effects in the decay channel}
\label{Subsect:Unitarity}

In the following we consider both decay modes of the $\eta$ meson, the charged decay \mbox{$\eta\rightarrow \pi^+\pi^-\pi^0$}, and the neutral decay $\eta\rightarrow 3\pi^0$. When comparing with experimental data it is important to have an accurate description of the phase space boundary, thus in the computation of the kinematical factors we use the physical pion masses. Elsewhere we assume the isospin limit and use 
$m_{i}  =  (2\,m_{\pi^{+}}+m_{\pi^{0}})/3 \equiv 
m_{\pi}$, {\it i.e.} the isospin averaged mass.

\begin{figure}[t]
\includegraphics[width=0.48\textwidth]{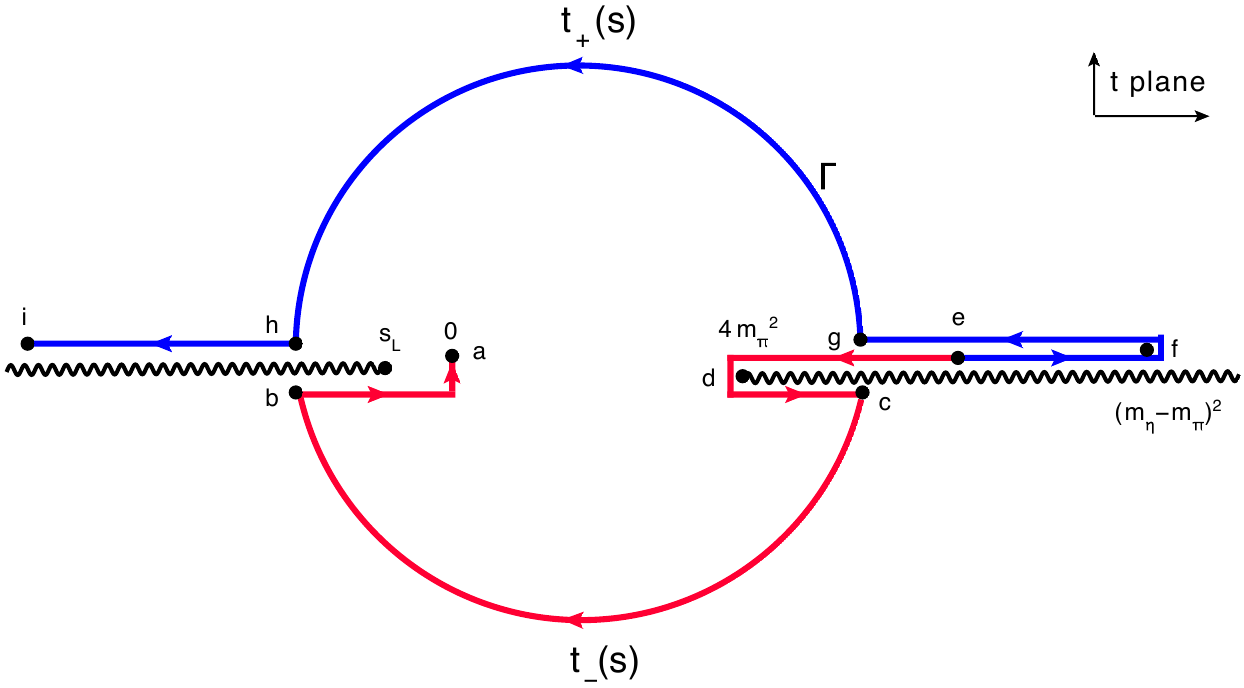}
\caption{Integration contour in the complex $t$ plane. The arrows indicate the direction of increasing $s$  in the interval from $4\,m_{\pi}^{2}$ to $\infty$. The points labeled $a$ through $i$  correspond to specific values of $s$, with  (a) $t_{-}( \infty)= 0$, (b) $t_{-}( (m_{\eta}+m_{\pi})^{2})=  m_{\pi} (m_{\pi}-m_{\eta})$,  (c) $t_{-}((m_{\eta}-m_{\pi})^{2} )=  m_{\pi}(m_{\eta}+m_{\pi})$, (d) $t_{-}( \frac{ m_{\eta}^{2} -m_{\pi}^{2} }{2})=4 m_{\pi}^{2}$, (e) $t_{\pm}( 4 m_{\pi}^{2})= \frac{ m_{\eta}^{2} -m_{\pi}^{2} }{2}$, (f) $t_{+}( m_{\pi}(m_{\eta}+m_{\pi}))= (m_{\eta}-m_{\pi})^{2}$,  (g) $t_{+}( (m_{\eta}-m_{\pi})^{2})=  m_{\pi}(m_{\eta}+m_{\pi})$, (h) $t_{+}((m_{\eta}+m_{\pi})^{2})=  m_{\pi}(m_{\pi}-m_{\eta})$,  and (i) $t_{+}(\infty)= -\infty$, respectively. \label{fig:Contour_1}}
\end{figure}

The model is defined by Eq.\,(\ref{Eq:AIL}) together with the elastic unitarity constraint for the right-hand discontinuity~\cite{Mandelstam:1960zz},
\begin{align}
\Delta a^{Right}_{I  L} (s) & \equiv \frac{1}{2\,i}\left(a^{Right}_{I  L} (s+ i \epsilon) - a^{Right}_{I  L} (s- i \epsilon)\right) \nonumber \\
&= f_{I  L}^{*}(s)\, \rho(s)\, (a^{Right}_{IL}(s) + a^{Left}_{I L}(s)), 
\end{align}
where $\rho(s) =\sqrt{1- 4\,m_{\pi}^{2}/s}$. The elastic $\pi\pi$ partial wave amplitudes are denoted by $f_{I  L}$ and normalized by \mbox{$\mbox{Im} (1/f_{I L}(s)) =-\rho(s)$}. Therefore, the amplitudes $a_{IL}(s)$ satisfy the relation, 
\begin{align}\label{dis}
\Delta a_{I  L} (s)  &= f_{I  L}^{*}(s) \rho(s) \bigg( a_{I L} (s)+\sum_{L'=0}^{L_{max}}\sum_{I'}\frac{2\,(2L'+1)}{K(s)/s}  \nonumber \\
&\quad \times\int_{t_{-}(s)}^{t_{+}(s)}dt\,P_{L}(z_{s})\,P_{L'}(z_{t}) C_{st}^{I  I'}  a_{I' L'} (t)   \bigg). 
\end{align}
The first term on the right-hand side of Eq.\,(\ref{dis})  represents the contribution from the direct $s$-channel, \mbox{$4 + \bar{3} \to 1 + 2$}, to the $s$-channel partial-wave  projection of the unitarity relation and it is illustrated in the diagram in Fig.~\ref{fig:dis}(a). The second term, illustrated in Fig.~\ref{fig:dis}(b), gives the contribution from the exchange contributions in the $t$-channel \mbox{$4 + \bar 1 \to  2 + 3$}, and $u$-channel \mbox{$4 + \bar 2 \to 1 +  3$}. In Eq.\,(\ref{dis}), using Eq.\,(\ref{zs}), we changed the integration over the $z_s$ to integration over $t$, 
\begin{equation}\label{dzs}
\int_{-1}^{1} \frac{dz_s}{2} \left ( \cdots \right ) = \int_{t_-(s)}^{t_+(s)} \frac{dt}{K(s)}\,s \left( \cdots \right ),
\end{equation}
with the integration limits \mbox{$t_{\pm}(s)$} corresponding  to \mbox{$z_s = \pm 1$},
\begin{equation} 
t_{\pm}(s) = \frac{m_\eta^2 + 3\,m_\pi^2 -s}{2} \pm \frac{K(s)}{2\,s}\,.
\end{equation} 
The Kacser function $K(s)$ is given by the product of the triangle functions and has the following determination \cite{PhysRev.132.2712,Kambor:1995yc}
\begin{align}
K(s)&=\left\lbrace
\begin{array}{ll}
+\kappa(s)\,,&\quad  4\,m_\pi^2\leq s\leq (m_\eta-m_\pi)^2 , \\
\,\,i\,\kappa(s)\,,&\quad (m_\eta-m_\pi)^2\leq s\leq (m_\eta+m_\pi)^2, \\
-\kappa(s)\,,&\quad (m_\eta+m_\pi)^2\leq s<+\infty ,
\end{array}
\right.\nonumber \\
\kappa(s)&=|\lambda(s,m_\eta^2,m_\pi^2)\,\lambda(s,m_\pi^2,m_\pi^2)|^{1/2}\,.
\end{align}
In the scattering region \mbox{$s \geq(m_\eta+m_\pi)^2$}  the integral in Eq. (\ref{dzs}) is well defined; however, when \mbox{$4\,m_\pi^2 \leq s <(m_\eta+m_\pi)^2$}, analytical continuation to the decay region is needed. For this  a positive infinitesimal imaginary part is added to the eta mass \cite{Gribov:1962fu,PhysRev.132.2712,Bronzan:1963kt}, which leads to the integration contour in the $t$-plane shown in Fig. \ref{fig:Contour_1}. It is worth noting that the contour avoids the unitary cut. 
Finally, the amplitudes $a_{IL}(s)$ are obtained by bootstrapping the dispersion reaction
\begin{equation} 
a_{IL}(s) = \frac{1}{\pi} \int_{4m_{\pi}^2}^\infty ds' \frac{\Delta a_{IL}(s')}{s' -s}, \label{dis2} 
\end{equation}
with $a_{IL}$ appearing on the right-hand side ({\it cf.} Eq.~(\ref{dis})) together with the input two-body scattering amplitudes, \mbox{$f_{IL}(s)$}.

As in the standard $N/D$ approach, the inhomogeneous part in Eq.~(\ref{dis}) can be accounted for writing $a_{IL}(s)$ as a product of $f_{IL}(s)$ times another function of $s$, whose discontinuity is given by the $s$-channel projection of the cross-channel amplitudes.  It is also convenient to remove any zeros of $f_{IL}(s)$, {\it e.g.} the Adler zero, since these are process dependent. Finally, the partial waves have kinematical singularities, which do not contribute to the discontinuity relation given by Eq.~(\ref{dis}). Thus, we write 
\begin{equation}\label{Eq:product}
a_{IL}(s) =  Z_L(s)  \mathcal{F}_{IL}(s) \,f_{IL}(s)  \,g_{IL}(s),
\end{equation}
where the first factor removes the kinematical singularities 
\begin{equation} 
Z_L(s) = \left[ \frac{ K(s)}{s/4- m_{\pi}^{2}}\right]^L
\end{equation} and 
the second factor removes zeros from the $\pi\pi$ amplitude, 
\begin{equation}\label{gredef}
\mathcal{F}_{IL}(s) =\left\{
\begin{array}{cl} (s-s_{\chi}^{(I)})/(s-s_{A}^{(I)}),& \quad  L=0~, \\ 
&\\
1,&\quad L>0~.
\end{array} 
\right.
\end{equation}
That is,  we assume $f_{IL}$ has zeros in the $S$-wave only.  Note that at leading order in  $\chi$PT, Adler zeros are located at \mbox{$s_{A}^{(0)}= m_{\pi}^{2}/2$} and \mbox{$s_{A}^{(2)}= 2\,m_{\pi}^{2}$} in the $\pi\pi$ $S$-wave  isoscalar and isotensor amplitudes, respectively,  and at \mbox{$s_{\chi}^{(0)}=4/3\,m_{\pi}^{2}$} for $\eta\to3\pi$. In the actual calculation we use as input the $\pi\pi$ amplitudes from the phenomenological analysis of \cite{GarciaMartin:2011cn} which have zeros at the same position as the leading order in $\chi$PT; when matching $\eta\rightarrow 3\pi$ with $\chi$PT we use  NLO calculation which places the zeros in  \mbox{$\eta\rightarrow 3\pi$}  at  \mbox{$s_{\chi}^{(0)}=1.25\,m_{\pi}^{2}$} and \mbox{$s_{\chi}^{(2)}=2.7\,m_{\pi}^{2}$} in the isoscalar and isotensor  channels, respectively.

Finally, it follows from Eq.~(\ref{dis}) and Eq.~(\ref{Eq:product}) that the function $g_{IL}$ has the discontinuity given by 
\begin{align}\label{gdis}
& \Delta g_{I  L} (s) =  -\,\theta(-s)\,\frac{\Delta f_{IL}(s)}{f_{IL}^{*}(s)}\,g_{I L} (s)  \\
& \quad +\theta(s-4 m_{\pi}^{2}) \sum_{L'=0}^{L_{max}}\sum_{I'}\frac{ 2\,(2L'+1)}{K(s)/s}    \frac{\rho(s) \,P_{L}(z_{s})}{ \mathcal{F}_{IL}(s) Z_L(s) }\nonumber \\
& \times \int_{t_{-}(s)}^{t_{+}(s)} d t\,\,P_{L'}(z_{t}) C_{st}^{I  I'}  Z_{L'}(t)  \mathcal{F}_{I'L'}(t)  f_{I'L'}(t)\,g_{I'L'}(t)\,.\nonumber 
\end{align}
The first  term on the left-hand side takes into account the left-hand cut of \mbox{$f_{IL}(s)$}; {\it i.e.} in addition to the unitary cut, $g_{IL}$ has a left-hand cut determined by $f_{IL}$ to guarantee that there is no dynamical left-hand cut in the amplitudes $a_{IL}$. The integrand in Eq.~(\ref{gdis}) is free from kinematical singularities in $t$ and the function $g_{IL}(s)$ satisfies 
\begin{equation} 
g_{IL}(s) = \frac{1}{\pi} \int_{-\infty}^\infty ds'\,\frac{\Delta g_{IL}(s')}{s' -s} .\label{gdisp} 
\end{equation}
Inserting Eq.~(\ref{gdis}) into Eq.~(\ref{gdisp}) we obtain a double integral equations for \mbox{$g_{IL}(s)$}, which can be reduced to a single integral equation by changing the order of dispersive integral (over $s$) and the angular projection (internal over $t$). The procedure, which we referred to earlier as the Pasquier inversion, was developed in \cite{Pasquier:1968kt,Aitchison:1978pw} and recently revisited in \cite{Guo:2014vya}. It leads to the following representation  
\begin{align}
g_{IL}(s)  = &  -\frac{1}{\pi} \int_{-\infty}^{ 0} ds' \frac{1}{s' - s}  \frac{\Delta f_{IL}(s)}{f_{IL}^{*}(s)}\,g_{IL}(s')   \nonumber \\
&+ \frac{1}{\pi} \int_{-\infty}^{(M-m_{\pi})^2}  dt\sum_{L'=0}^{L_{max}}\sum_{I'} \mathcal{K}_{ I L, I' L'}(s,t)  \nonumber \\
& \quad \quad \quad \quad  \quad \quad  \quad    \times   C_{st}^{I  I'}   f_{I' L'}(t)\,g_{I'L'}(t)   , \label{pasqg}
\end{align} 
where the kernel function \mbox{$\mathcal{K}_{IL, I' L'}(s,t)$} is given explicitly in Appendix \ref{App:B}. The left-hand cut contribution to $g_{IL}(s)$ is largely unknown. Since we are primarily   interested in the physical decay region we therefore parametrize contributions to $g_{IL}$ from integration over $s<0$.  In the simplest approximation these are reduced to a constant. A more elaborated representation could, for example, involve a conformal  map of the $s$-plane cut along the negative real axis onto a unit circle ~\cite{Yndurain:2002ud}. However, in the analysis of the data we find the simple approximation to be sufficient: 
\begin{align}\label{gapprox}
&g_{IL}(s)  =   g_{IL}(s_0) + \frac{1}{\pi} \int_{0}^{(M-m_{\pi})^2} dt \sum_{L'=0}^{L_{max}}\sum_{I'}  
C_{st}^{I  I'}  \nonumber \\
 &\quad \quad \times \big(\mathcal{K}_{IL,I'L'}(s,t) -\mathcal{K}_{IL,I'L'}(s_{0},t)\big)  f_{I' L'}(t)\,g_{I'L'}(t)\,.
\end{align}
This equation can now be solved using standard matrix inversion methods with the subtraction constants $g_{IL}(s_0)$ as fitting parameters. The subtraction point is arbitrary and we choose it to coincide with the Adler zero of the LO $\chi$PT \mbox{$s_0= 4/3\,m_\pi^2$}. After solving the integral equation for $g_{IL}(s)$, we compute \mbox{$a_{IL}(s)$} from Eq.~(\ref{Eq:product}). Finally, to compare with the experimental data we convert the isospin amplitudes to the charge amplitude, \mbox{$A^{C}(s,t,u)$} for the $\eta \to \pi^+\pi^-\pi^0$ and \mbox{$A^{N}(s,t,u)$} for the neutral case. These are given by Eq. (\ref{decayamp}), 
\begin{align} 
& A^{C}(s,t,u) =     \sum_{L=0}^{L_{max}}\frac{(2L+1)}{2} \bigg [ \frac{2}{3}\,P_{L}(z_{s})  \left ( \, a_{0 L}  (s ) - a_{2 L} (s) \right)  \nonumber \\
& + P_{L}(z_{t})  \left ( \,a_{1 L} (t) +  a_{2 L} (t) \right)-P_{L}(z_{u})  \left ( \,a_{1 L} (u) - a_{2 L} (u) \right)  \bigg ]\,,\nonumber \\
& A^{N}(s,t,u)=\sum_{L=0}^{L_{max}}\frac{(2L+1)}{3} \bigg[ \,P_{L}(z_{s})\left(\,a_{0 L} (s) +2a_{2 L} (s) \right) \nonumber \\
& \quad\quad\quad\quad\quad\quad\quad\quad\quad + (s \to t) + (s\to u) \bigg]\,. \label{ampn}
\end{align}

\section{Numerical results} 
\label{SectionIII}
In this section we present our results for the decays \mbox{$\eta\rightarrow \pi^+\pi^-\pi^0$} and \mbox{$\eta\rightarrow 3\pi^0$}. We study the systematic uncertainties of the model by using different sets of partial waves, {\it i.e.} varying  $L_{max}$ and maximal isospin. We have found that  partial waves  with  (\mbox{$L\geq 2 $}) are negligible in the physical decay region, \mbox{$4\,m_\pi^2\leq s \leq (m_{\eta}-m_{\pi})^2$}. As input we use two-pion scattering amplitudes from the analysis  of \cite{GarciaMartin:2011cn}. The parameters of the fit are the subtraction constants, $g_{IL}(s_0)$, for each contributing  partial wave.  Our aim is to fix these by fitting \mbox{$\eta\rightarrow \pi^+\pi^-\pi^0$} decay using the high statistic WASA-at-COSY data \cite{Adlarson:2014aks} and by matching to NLO $\chi$PT \cite{Gasser:1984pr}. The results for the \mbox{$\eta\rightarrow 3\pi^0$} decay mode will then constitute a prediction, which we compare with the Dalitz plot distribution from \cite{Prakhov:2008ff}. We investigate the role of cross-channel exchanges, {\it a.k.a.} final-state interactions in the decay region, by performing two analyses. In the first, we do not include cross-channel effects and approximate \mbox{$g_{IL}(s)$} in Eq.\,(\ref{gapprox}) by a constant, setting \mbox{$g_{IL}(s)=g_{IL}(s_0)$}. It corresponds to a traditional isobar model, but with a fully incorporated two-pion interaction. In the second, we include cross-channel rescattering effects and solve Eq.\,(\ref{gapprox}). In the following we refer to the two cases as ``two-body" and ``three-body", respectively.  

\begin{table*}[t]
\centering
\caption{Results of two-body and three-body fits for different wave sets. 
\label{tab:par}}
\renewcommand{\arraystretch}{1.8}
\begin{tabular*}{\textwidth}{@{\extracolsep{\fill}}lcccl@{}}%\columnwidth}
\hline\hline
&    $g_{00}(s_0)/g_{00}^{(2b)}$ &  $g_{20}(s_0)/g_{00}^{(2b)}$ & $g_{11}(s_0)/g_{00}^{(2b)}$ & \mbox{$\chi^{2}/d.o.f.$}\\
\hline
$(I,L)=(0,0)$ &&&&\\
\hline
two-body & $1.000 \pm0.002$& -- & --& $2.2$\\
three-body& $1.062 \pm 0.002$ & --& --& $15$ \\
\hline
$(I,L)=(0,0),\,(2,0)$ &&&&\\
\hline
two-body  &  $1.000 \pm 0.003$ &    $0.04\pm0.01$ & -- & $1.69$  \\
three-body&  $1.138\pm0.003$ &$0.29\pm 0.01$& -- & $1.67$ \\
\hline
$(I,L)=(0,0),\,(1,1)$ &&&&\\
\hline
two-body  & $1.000 \pm 0.002$        & --          &  $0.058 \pm 0.009$& $1.45$ \\
three-body&$1.043 \pm 0.005$& -- & $0.233 \pm 0.009$& $0.95$ (Set 1)\\
\hline
$(I,L)=(0,0),\,(2,0),\,(1,1)$ &&&&\\
\hline
two-body  & $1.00 \pm 0.02$& $-0.26\pm 0.05$ & $0.38\pm 0.07$ & $0.94$ \\
three-body &$1.19 \pm 0.01$        & $0.14\pm 0.03$ & $0.28 \pm 0.04$ & $0.90$ (Set 2)\\
\hline\hline
\end{tabular*}
\end{table*}

\begin{table*}[t]
\centering
\caption{Dalitz plot parameters for \mbox{$\eta \to \pi^{+} \pi^{-} \pi^{0}$}. Set 1 and Set 2 correspond to  \mbox{$(I,L)=(0,0),\,(1,1)$} and \mbox{$(I,L)=(0,0),\,(2,0),\,(1,1)$} cases respectively (see Table \ref{tab:par}). \label{tab:xy}}
\renewcommand{\arraystretch}{1.8}
%\fontsize{8}{2}
\begin{tabular*}{\textwidth}{@{\extracolsep{\fill}}llllll@{}}
\hline\hline
 &  $a$&    $b$ &  $d$ & $f$ & $g$ \\
\hline
%Experiment   & &&&  \\
%
WASA-at-COSY \cite{Adlarson:2014aks}      & $-1.144\pm0.018$ &      $0.219\pm0.019\pm0.037$ 
%&  $-0.007\pm 0.009$  
& $0.086\pm0.018\pm0.018$ & $0.115\pm0.037$  & -- \\
KLOE  \cite{Ambrosino:2008ht}      & $-1.090\pm0.005^{+0.008}_{-0.019}$ &      $0.124\pm0.006\pm0.010 $  & $0.057\pm0.006^{+0.007}_{-0.016}$ & $ 0.14\pm0.01\pm0.02$ & -- \\
CBarrel  \cite{Abele:1998yj}    & $-1.22\pm0.07 $ &      $0.22\pm0.11$  & $0.06\pm0.04\,$(fixed) & -- &-- \\

Layter {\it et al.}  \cite{Layter:1973ti}      & $-1.080\pm0.014 $ &      $0.03\pm0.03$  & $0.05\pm0.03$  & -- & --\\

Gormley {\it et al.}  \cite{Gormley:1970qz}      & $-1.17\pm0.02 $ &      $0.21\pm0.03$  & $0.06\pm0.04$& -- & -- \\
\hline
Theory &&&&\\\hline
Set 1 &  $-1.116\pm0.030 $&     $0.188 \pm 0.010  $  &$0.047 \pm 0.005 $ & $0.093 \pm 0.004 $ & $-0.020\pm 0.006$ \\
Set 2 &  $-1.117\pm0.035 $&     $0.188 \pm 0.014  $  &$0.079 \pm 0.003 $ & $0.090 \pm 0.003 $ & $-0.063\pm 0.012$ \\
\hline\hline
NLO \cite{Gasser:1984pr}&   $-1.371$   & $0.452$ & $0.053$  & $0.027$ & -- \\
NNLO \cite{Bijnens:2007pr}&   $-1.271 \pm 0.075$   & $0.394\pm0.102$ & $0.055\pm0.057$  & $0.025\pm0.160$  & -- \\
Kambor {\it et al.} \cite{Kambor:1995yc}      & $-1.16 $ &      $0.24 ... 0.26$  & $0.09 ... 0.10$  & -- & -- \\
NREFT \cite{Schneider:2010hs}     & $-1.213\pm0.014 $ &      $0.308\pm0.023$  & $0.050\pm0.003$  & $0.083\pm0.019$ & $-0.039\pm0.002$ \\
\hline\hline
\end{tabular*}
\end{table*}

\subsection{Fitting WASA-at-COSY data} 
\subsubsection{\mbox{$\eta\rightarrow \pi^+\pi^-\pi^0$}}
In this subsection we summarize the results of the fit to the recent WASA-at-COSY data on \mbox{$\eta \to \pi^{+} \pi^{-} \pi^{0}$} \cite{Adlarson:2014aks}, where binned Dalitz plot is given. 
Up to a normalization factor, the Dalitz plot distribution is given by the amplitude squared, 
\begin{equation}
\frac{d^2\Gamma}{ds\,dt}\propto |A(s,t)|^2. 
\end{equation}
It is convenient to express the amplitude in terms of two independent, dimensionless variables \mbox{$(x,y)$} which are linearly related to the Mandelstam variables by 
\begin{align}\label{defXY}
x &= \frac{\sqrt{3}}{2\,m_\eta\,Q_{c}}\,(t-u)\,, \nonumber\\
y &= \frac{3}{2\,m_\eta\,Q_{c}} \left ((m_\eta-m_{\pi^{0}})^{2} -s \right ) -1\,,
\end{align}
where \mbox{$Q_{c} = m_\eta-2\,m_{\pi}^+-m_\pi^0$} (for the neutral decay we use \mbox{$Q_{n} = m_\eta-3\,m_\pi^0$}). A general property of these variables is that the physical region of the Dalitz plot lies inside the unit circle \mbox{$x^2+y^2\leq1$} centered at \mbox{$x=y=0$}. We fit our model to the data \cite{Adlarson:2014aks} by minimizing   the $\chi^{2}$ defined by 
\begin{align}
\chi^{2} = \sum^{N}_{\text{bins}}  \left ( \frac{|A|_{\mbox{data}}^{2} - |A^{C}\left(\{g_{IL}(s_{0})\}\right)|^{2} }{\Delta|A|_{\mbox{data}}^{2} } \right)^{2}\,, \label{c2} 
\end{align}
over the set of subtraction constants, $g_{IL}(s_{0})$. In Eq.~(\ref{c2}), \mbox{$|A|_{\mbox{data}}$} is the acceptance-corrected number of events in each of the $N=59$, \mbox{$\Delta x=\Delta y=0.2$} wide mass bins. The data is normalized to unity at $x=y=0$ and \mbox{$\Delta|A|_{\mbox{data}}$} is the statistical uncertainty. Note, that since Eq. (\ref{gapprox}) is linear in $g_{IL}$, the parameter $g_{00}(s_0)$ can be factored out and fixed by the overall normalization. Since normalization of the data is arbitrary the absolute value of $g_{00}(s_0)$ is irrelevant. Therefore, in Table \ref{tab:par}, which summarized fit results, when presenting results of two-body fits we quote $(g^{2b}_{IL}(s_0) \pm \Delta g^{2b}_{IL}(s_0))/g^{2b}_{00}(s_0)$.  When presenting results of three-body fit we quote $(g^{3b}_{IL}(s_0) \pm \Delta g^{3b}_{IL}(s_0))/g^{2b}_{00}(s_0)$, where $g^{2b}_{00}(s_0)$ is the central value obtained in the two-body fit with the same number of partial waves. We do the latter to illustrate the relative change in normalization between two- and three-body fits.

\begin{figure*}[t]
\includegraphics*[width=0.49\textwidth]{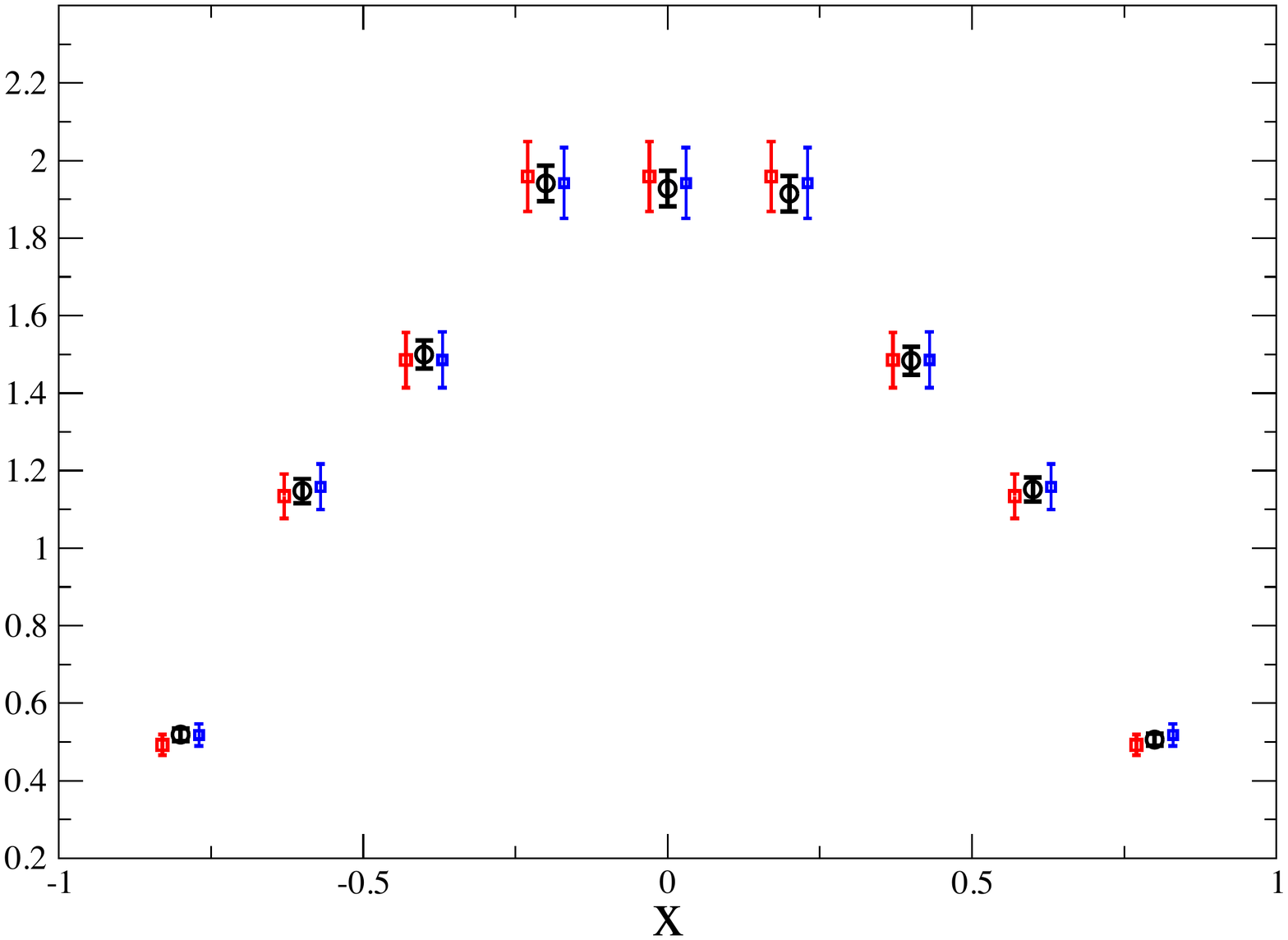}
\includegraphics*[width=0.49\textwidth]{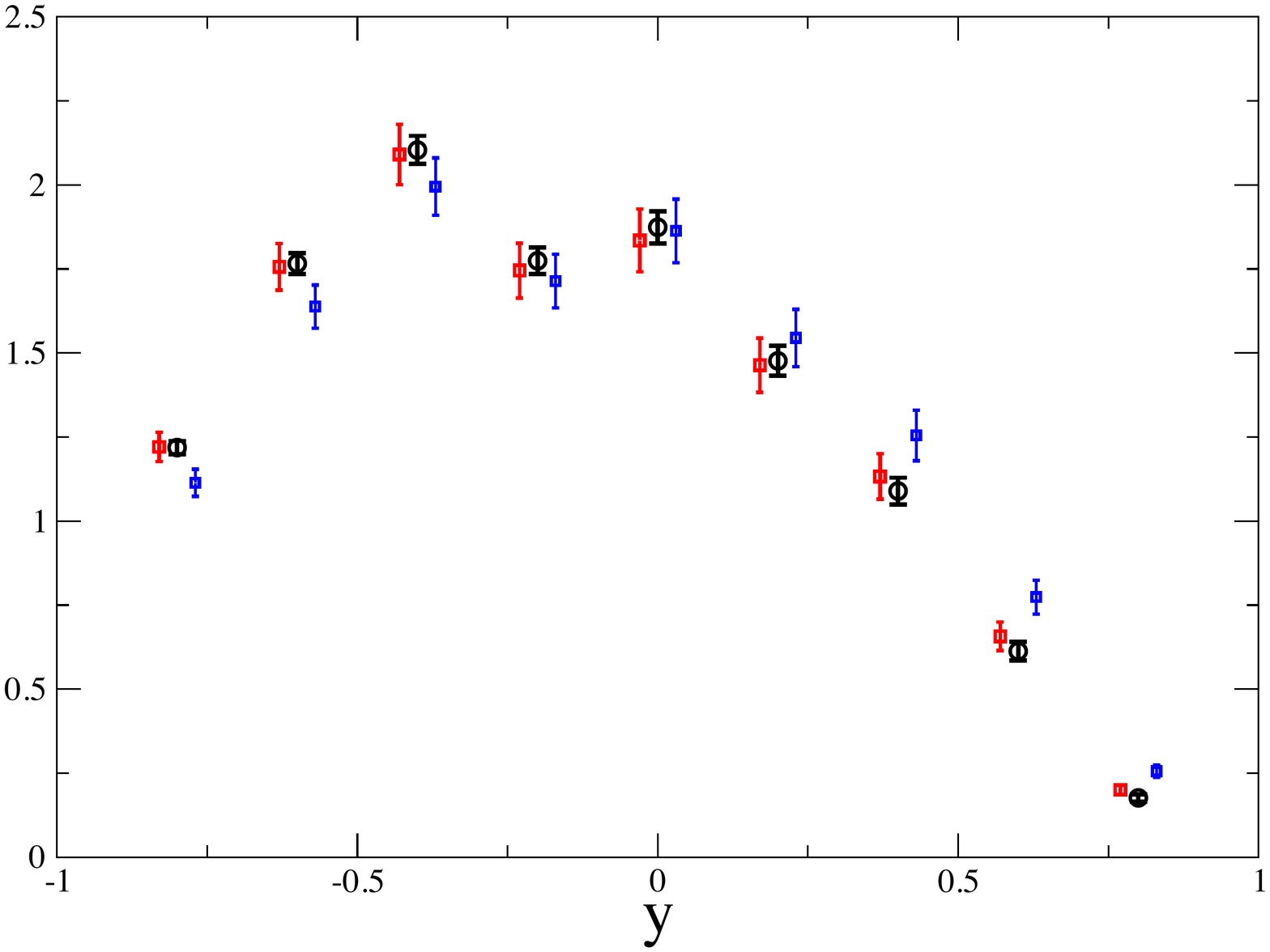}
\includegraphics*[width=0.49\textwidth]{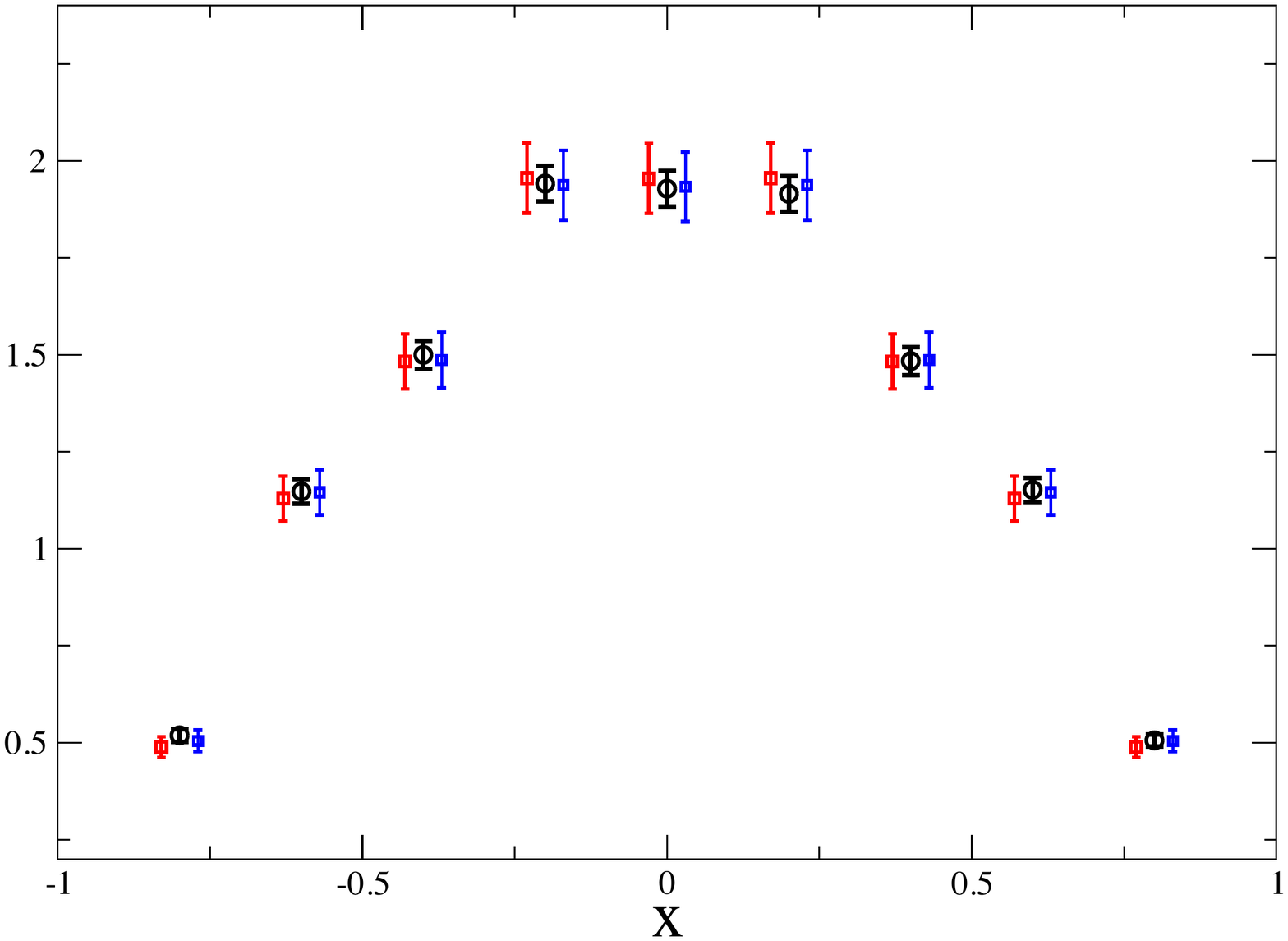}
\includegraphics*[width=0.49\textwidth]{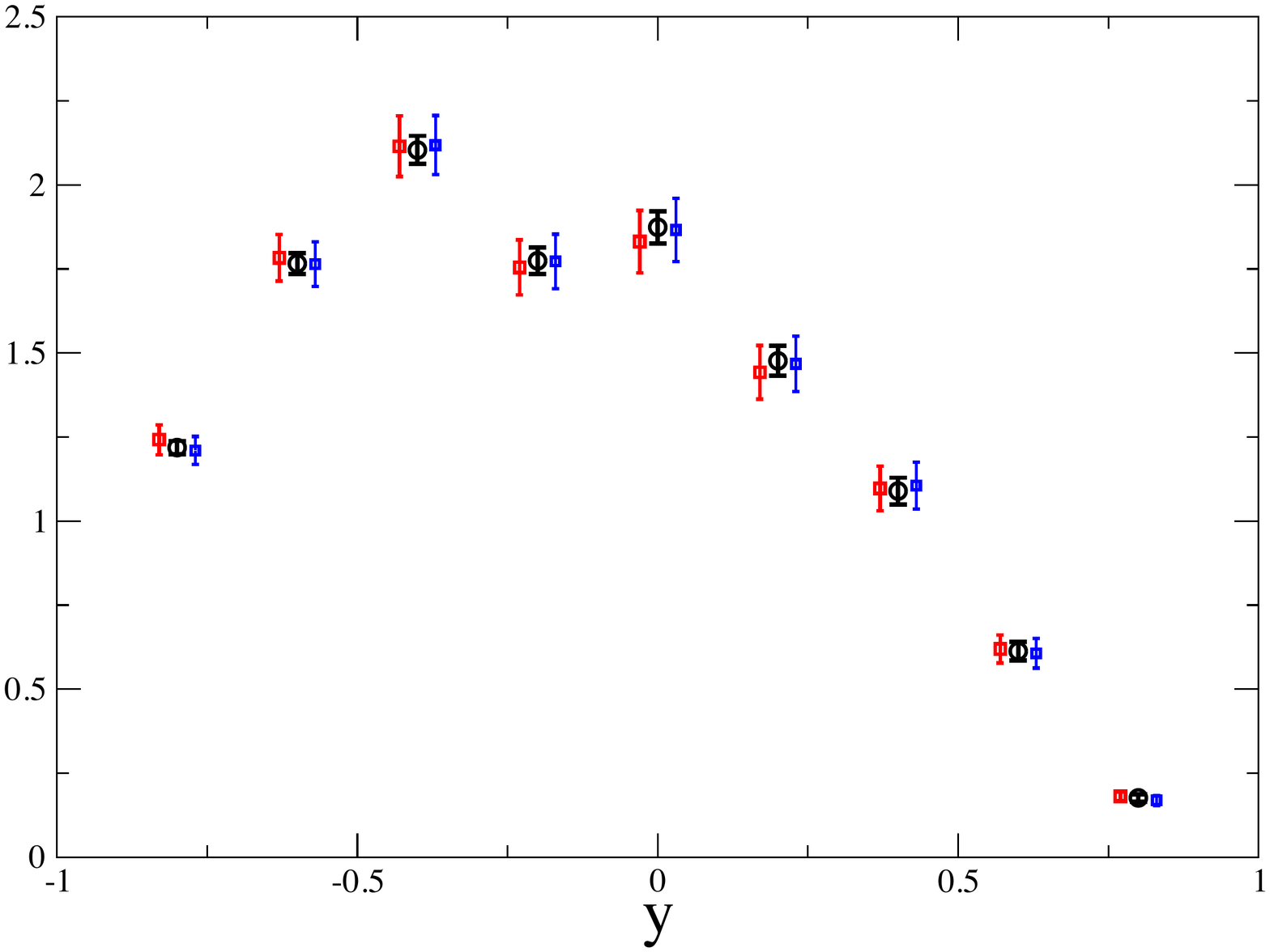}
\includegraphics*[keepaspectratio,width=0.49\textwidth]{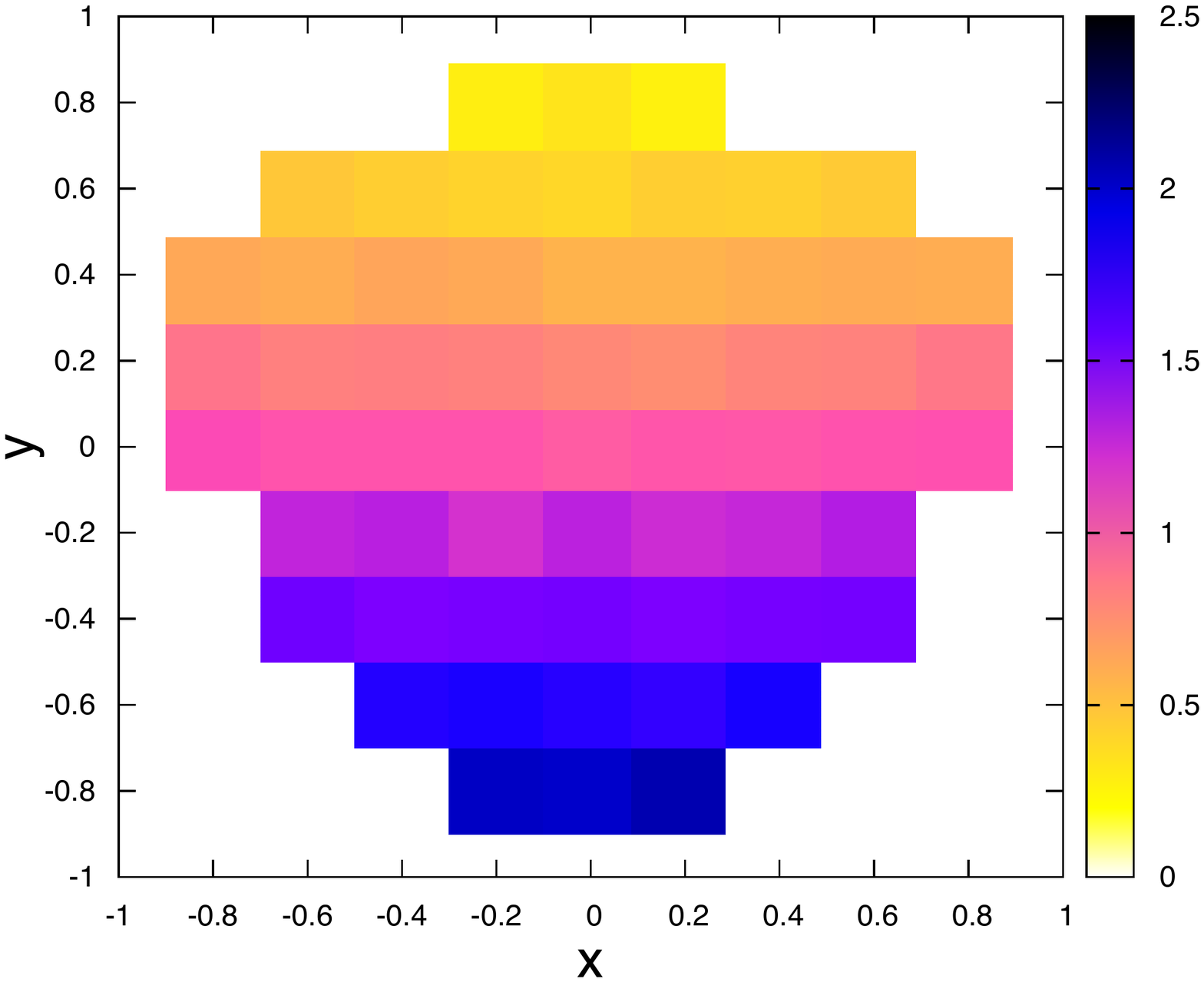}
\includegraphics[keepaspectratio,width=0.49\textwidth]{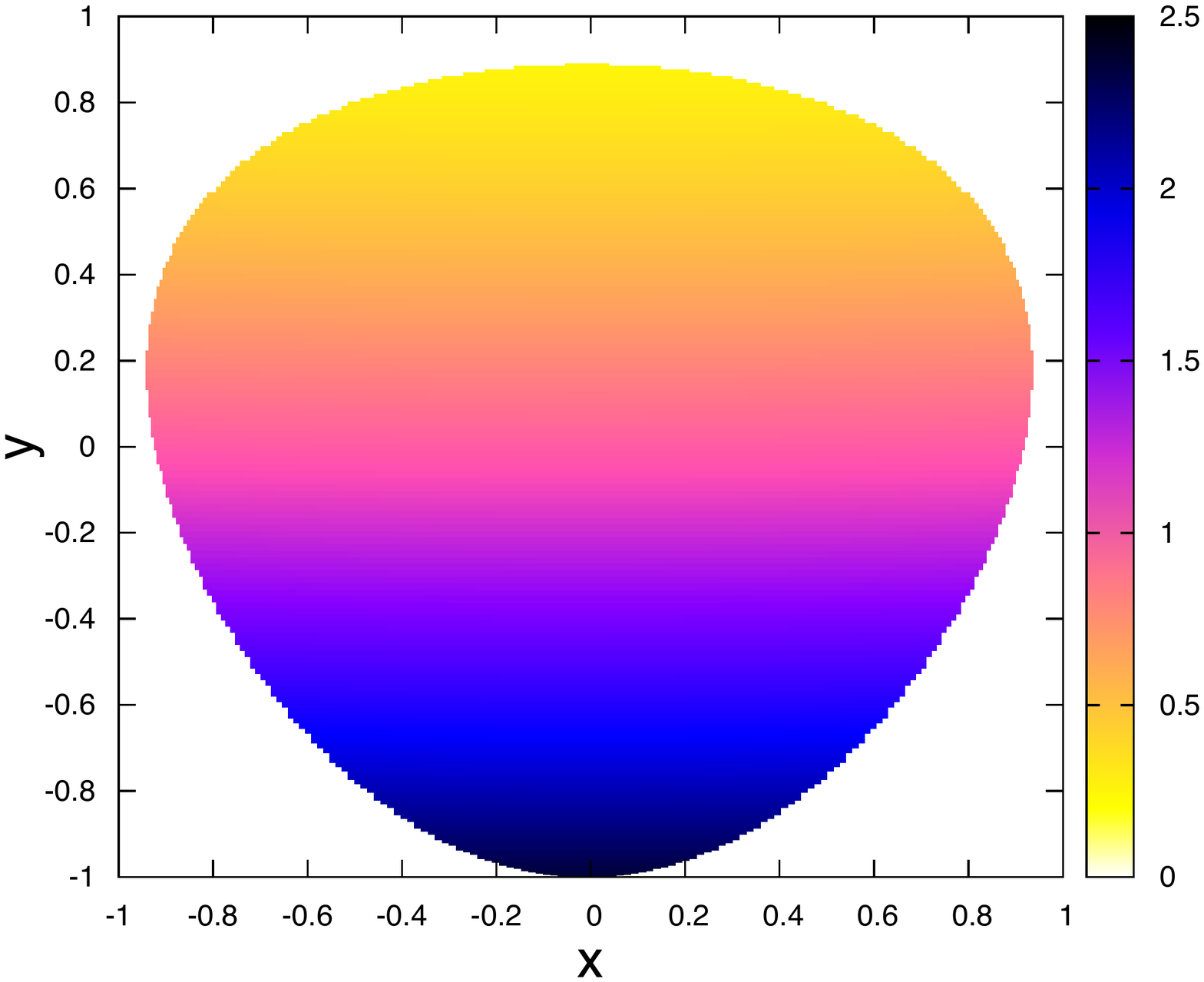}
\caption{Upper and middle panels are the $x$- and $y$-projection plots.  Black circles are the data. Red squares and blue squares represent results of the  two-body and three-body fits, respectively. The fits are performed on the  Dalitz distribution \cite{Adlarson:2014aks} sown in the bottom left panel using a single,  \mbox{$(I,L)=(0,0)$}  wave (upper panels) and two waves, \mbox{$(I,L)=(0,0),\,(1,1)$} (central panels). For better visualization fit results are shifted horizontally  (three-body to right and two-body to left) from the experimental points. The bottom right panel is the Dalitz distribution from the three-body fit with \mbox{$(I,L)=(0,0),\,(1,1)$} waves. 
\label{fig:proj1}}
\end{figure*}

In the first fit we use a single, scalar-isoscalar, $a_{00}$ partial wave. In this case, the model gives a parameter free prediction for the event distribution. We observe that the $(I,L)=(0,0)$ amplitude provides the dominant contribution that covers approximately $90\%$ of the Dalitz plot. The calculated \mbox{$\chi^{2}/d.o.f.$} for the two-body and three-body cases  are $2.2$ and $15$, respectively.  In Fig.~\ref{fig:proj1} (upper panels) we compare our results and the data  projected onto the $x$ and $y$ axes. The error bars associated with the model originate from the uncertainties in the pion-pion amplitude $f_{IL}$ \cite{GarciaMartin:2011cn} and  from the statistical error in fitting the overall normalization.

In the next step, we add the isospin-2 $S$-wave. In this case we fit two parameters, one gives the overall normalization and the other contributes to a modification of the shape of the Dalitz plot. The resulting parameters and \mbox{$\chi^2/d.o.f$} are given in Table \ref{tab:par}. In both, the two- and three-body fits we find that the model slightly underestimates the data. The inclusion of the second \mbox{$(I,L)=(2,0)$} wave significantly improves $\chi^2$ and also drastically reduces the difference in the fit quality between the two- and three-body cases pertinent in the fit with the single $(I,L)=0$ wave.

In the spirit of keeping the number of free parameters as low as possible, we considered another set of two waves,  \mbox{$(I,L)=(0,0),\,(1,1)$}, before taking into account a complete sum of $S$ and $P$ waves. In this case there is also one parameter that affects the shape of the Dalitz distribution and we find \mbox{$\chi^2/d.o.f=1.45$} and \mbox{$\chi^2/d.o.f=0.95$} in the two-body and three-body fits, respectively. Hence, it seems that the data favor the isovector $P$-wave contribution over the isospin-2 $S$-wave. The results of the fit are shown in Fig. \ref{fig:proj1}.

We now turn to the case when a complete set of $S$ and $P$ waves is incorporated, i.e. \mbox{$(I,L)=(0,0),\,(2,0),\,(1,1)$}. The two- and three-body fits result in a comparable \mbox{$\chi^2/d.o.f$} around $0.9$.

It is instructive to compare the results of the three-body fits.  In the fit with a single $(I,L)=(0,0)$ amplitude, the three-body fit converges poorly indicating importance of higher partial waves that are brought in by the cross-channel exchanges. Thus apparent convergence of the two-body fit in this case is deceptive. With any combination of higher partial waves all calculated three-body \mbox{$\chi^2/d.o.f$} are quite similar to the two-body fits, except for the case when only \mbox{$(I,L)=(0,0),\,(1,1)$} amplitudes were considered.

Often, an effective range expansion of the Dalitz plot near  $x=y=0$ is used to parametrize the $\eta$ decay distribution. For the charged decay it leads to 
\begin{align}\label{Dalitzpar1}
\frac{|A^{C}(x,y)|^{2}}{|A^{C}(0,0)|^{2} } & =1+ a\,y + b\,y^{2}+ c\,x+ d\,x^{2 }\nonumber \\   
&\quad \quad  +\, e\,xy  + f\,y^{3} + g\,x^{2} y + \cdots\,.
\end{align} 
The charge conjugation symmetry,  \mbox{$x\rightarrow -x$}  requires terms odd in $x$ to vanish, i.e. \mbox{$c=e=0$}. In Table ~\ref{tab:xy} we give the Dalitz plot parameters from our three-body fits based on the \mbox{$(I,L)=(0,0),\,(1,1)$} (set 1) and \mbox{$(I,L)=(0,0),\,(2,0),\,(1,1)$} (set 2) wave sets. For comparison we quote the results of next-to-leading-order (NLO) and next-to-next-to leading  order (NNLO) of $\chi$PT \cite{Gasser:1984pr,Bijnens:2007pr}, the dispersive analysis from  \cite{Kambor:1995yc}, NREFT \cite{Schneider:2010hs} and alternative dispersive approach \cite{Kampf:2011wr}. We also include Dalitz parameters extracted from direct fits to the experimental data   \cite{Adlarson:2014aks,Ambrosino:2008ht,Abele:1998yj,Layter:1973ti,Gormley:1970qz}.  The most recent analyses where performed by the WASA-at-COSY \cite{Adlarson:2014aks} and KLOE \cite{Ambrosino:2008ht} collaborations. As expected, our Dalitz plot parameters are consistent with the WASA-at-COSY parameters within the error bars. We also observe that central values of the fit tend toward the KLOE results.

\subsubsection{\mbox{$\eta\rightarrow 3\pi^0$}}

The results obtained in the  charged mode  can be used to predict the Dalitz plot parameters for the neutral channel. The Dalitz parameters are defined as coefficients in the expansion around the center of the Dalitz plot using the polar coordinates \mbox{$x=\sqrt{z} \cos \phi$} and \mbox{$y=\sqrt{z} \sin \phi$} in Eq.\,(\ref{defXY})
\begin{align} 
& \frac{|A^{N}(z,\phi)|^{2}}{|A^{N}(0,0)|^{2}}= 1+ 2\,\alpha\,z + 2\,\beta\,z^{3/2}\,\sin 3 \phi  + \cdots\,.
\end{align}
The slope parameter $\alpha$ has been extracted from several experiments, while to the best of our knowledge, there is no  determination of  $\beta$  or higher moments. In Table \ref{tab:alpha} we compare our findings with the experimental measurements and other theoretical predictions. The average of experimental results compiled by the PDG is \mbox{$\alpha=-0.0317\pm 0.0016$} \cite{PDG-2012}.

\begin{table}[tbp]
\centering
\caption{Dalitz plot parameters for \mbox{$\eta \to 3 \pi^{0}$}.  Set 1 and Set 2 correspond to  \mbox{$(I,L)=(0,0),\,(1,1)$} and \mbox{$(I,L)=(0,0),\,(2,0),\,(1,1)$} cases respectively (see Table \ref{tab:par}). \label{tab:alpha}}
\renewcommand{\arraystretch}{1.8}
%\fontsize{8}{2}
\begin{tabular*}{\columnwidth}{@{\extracolsep{\fill}}llc@{}}
\hline\hline
           & \multicolumn{1}{c}{$\alpha$}&    $\beta$\\
\hline
GAMS-2000 \cite{Alde:1984wj}   & $-0.022\pm 0.023$ &-- \\
Crystal Barrel, LEAR \cite{Abele:1998yi}   & $-0.052\pm 0.020$ &-- \\
Crystal Ball, BNL \cite{Tippens:2001fm}   & $-0.031\pm0.004$ &-- \\
SND \cite{Achasov:2001xi}   & $-0.010\pm 0.023$ &-- \\
CELSIUS-WASA \cite{Bashkanov:2007aa}   & $-0.026\pm 0.014$&--  \\
WASA-at-COSY \cite{Adolph:2008vn}  & $-0.027\pm0.009$ &-- \\
MAMI-B \cite{Unverzagt:2008ny}   & $-0.032\pm 0.004$&--  \\
MAMI-C \cite{Prakhov:2008ff}   & $-0.032\pm 0.003$&--  \\
KLOE \cite{Ambrosinod:2010mj}  & $-0.0301\pm 0.0050$&--  \\
PDG average \cite{PDG-2012}&$-0.0317\pm 0.0016$  & --\\
\hline
Theory &&\\
\hline
Set 1 & $-0.023  \pm0.004 $ & $-0.000 \pm 0.002$ \\
Set 2 & $- 0.020\pm 0.004$ & $-0.001 \pm 0.003$\\
\hline
NLO \cite{Gasser:1984pr}    & $+0.013$  &  -- \\
NNLO \cite{Bijnens:2007pr}     & $+0.013\pm 0.032$  &  -- \\
Kambor \textit{et al.} \cite{Kambor:1995yc}   & $-0.007\,... -0.014$ & --  \\
NREFT \cite{Schneider:2010hs}     & $-0.025\pm 0.005 $ & $-0.004\pm 0.001$ \\
Kampf \textit{et al.} \cite{Kampf:2011wr}    & $-0.044\pm 0.004 $ & -- \\
\hline\hline
\end{tabular*}
\end{table}

\begin{figure}[t]
\includegraphics[width=0.45\textwidth]{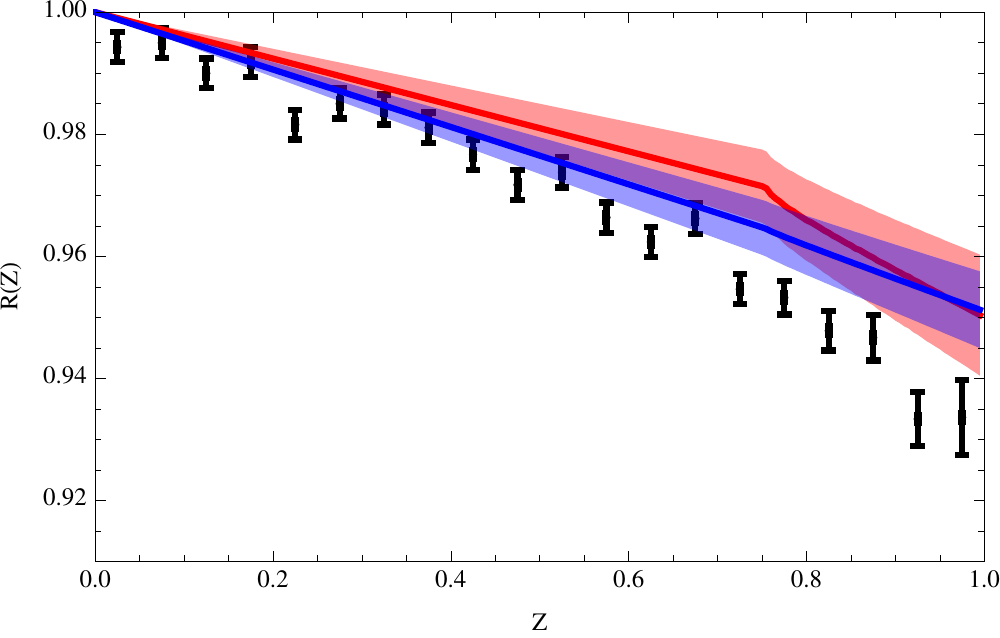}
\caption{Comparison of $R(z)$ plot from \cite{Prakhov:2008ff} (black points) with our predictions from Table~\ref{tab:alpha} that correspond to Set 1 (blue  band) and Set 2 (red band).
\label{fig:zplot}}
\end{figure}

As in the case of the charged mode,  our results obtained with the the two sets of waves are quite similar.  The predicted slope parameter is  \mbox{$\alpha(\text{Set 1})=-0.023$} and \mbox{$\alpha(\text{Set 2})=-0.020$}. Even though both sets describe the charged data well, the predicted slope parameter in the neutral case is above the PDG value. As shown in \cite{Bijnens:2007pr,Schneider:2010hs} the Dalitz plot parameters of the neutral and charged decays are related by 
\begin{align}
\alpha&=\frac{Q_n^2}{4\,Q_c^2}\,\left(d+b-\frac{1}{4}\,a^2-\text{Im}(\bar{a})^2\right)\nonumber\\
&\leq \frac{Q_n^2}{4\,Q_c^2}\,\left(d+b-\frac{1}{4}\,a^2\right),\label{Eq:alpha}
\end{align}
where the factors $Q_c$, $Q_n$ were defined below Eq.~(\ref{defXY}). Note that we only take $Q_c\neq Q_n$ in the overall normalization while we use $Q_c = Q_n$ when solving dispersion relations for the partial wave amplitudes. Here, the complex parameters $\bar{a}$ is the coefficient of the linear term in the  expansion of the charged amplitude \mbox{$A^C(x,y)$}, 
\begin{equation}
A^C(x,y)\propto 1+\bar{a}\,y+...
\end{equation}
Using the Dalitz plot parameters from WASA-at-COSY  and KLOE collaborations one finds 
\begin{equation}\label{Eq:alphaLimit}
\alpha^{\text{WASA}}\leq -0.006\,,\quad \quad
\alpha^{\text{KLOE}}\leq-0.033\,.
\end{equation}
The large difference in the upper limits is due to the difference in the $b$ parameter which differs by a factor of two between the two data sets. As pointed out in \cite{Schneider:2010hs} the value for $\text{Im}(\bar{a})$ can be sizable due to \mbox{$\pi\pi$} final state interactions. Our results confirm this finding and we obtain  \mbox{$\mbox{Im}(\bar{a}) = - 0.18 \pm 0.03$}. Nevertheless, since \mbox{$(\alpha^{\text{WASA}})_{max}=-0.006$} is quite large the $\text{Im}(\bar{a})$ term alone can not be responsible for lowering $\alpha$ to the PDG value. Once the KLOE data become available \cite{Balkestahl:2015aka} it would be very interesting to perform a combined fit of the WASA-at-COSY and KLOE measurements.

The neutral channel does not depend on the $P$-wave amplitude contributing to the charged decay mode and it contains only even partial waves. Unfortunately, using the charge mode  we could not find sensitivity to the $D$-wave  which was omitted from the Table ~\ref{tab:par}. Finally in Fig \ref{fig:zplot}, we compare  our results with the recent MAMI-C measurement  \cite{Prakhov:2008ff}. The $R(z)$ function is determined as
\begin{align}
R(z) = \frac{\int_{0}^{2\pi}d \phi  \, \theta(\varphi(s,t,u)) \, \frac{ |A^{N}(z,\phi)|^{2} }{  |A^{N}(0,0)|^{2}} }{\int_{0}^{2\pi}d \phi \, \theta (\varphi(s,t,u))}\,,\label{Rz}
\end{align}
where
\begin{equation}
\varphi(s,t,u)=s\,t\,u- m_{\pi^0}^{2} \,(m_\eta^{2}-m_{\pi^0}^{2})^{2}=0
\end{equation}
defines the boundary of the Dalitz plot distribution and $\theta(x)$ is the step function. We observe that a cusp around \mbox{$z\simeq0.765$} appears in $R(z)$ for nonzero $\beta$.  This is a kinematical effect which reflects the fact that for larger $z$ the phase space distribution in the Dalitz plot is no longer circular. We find our results for Sets 1 and 2 provide a satisfactory agreement with the data.

\subsection{Matching to $\chi$PT and the $Q$-value}
We remind that the data in \cite{Adlarson:2014aks} were normalized to the center of the Dalitz plot and therefore our model only predicts the  Dalitz plot distributions for the charged and neutral decays. The overall normalization can be fixed by comparing the experimental decay widths with the phase space integral over the corresponding squared amplitudes,
\begin{equation}
\Gamma  = N \int dx\,dy\,\frac{|A(x,y)|^{2}}{|A(0,0)|^{2}}\,,
\end{equation}
with the boundaries of the integral determined by the phase space. We emphasize that the quantity $Q^2$ defined in Eq.~(\ref{Eq:Q}) enters into the normalization constant $N$. In order to determine $Q^2$ one has to match the model, dispersive amplitude, with $\chi$PT where $Q^2$ is defined.

\begin{figure*}[t]
\includegraphics*[width=0.49\textwidth]{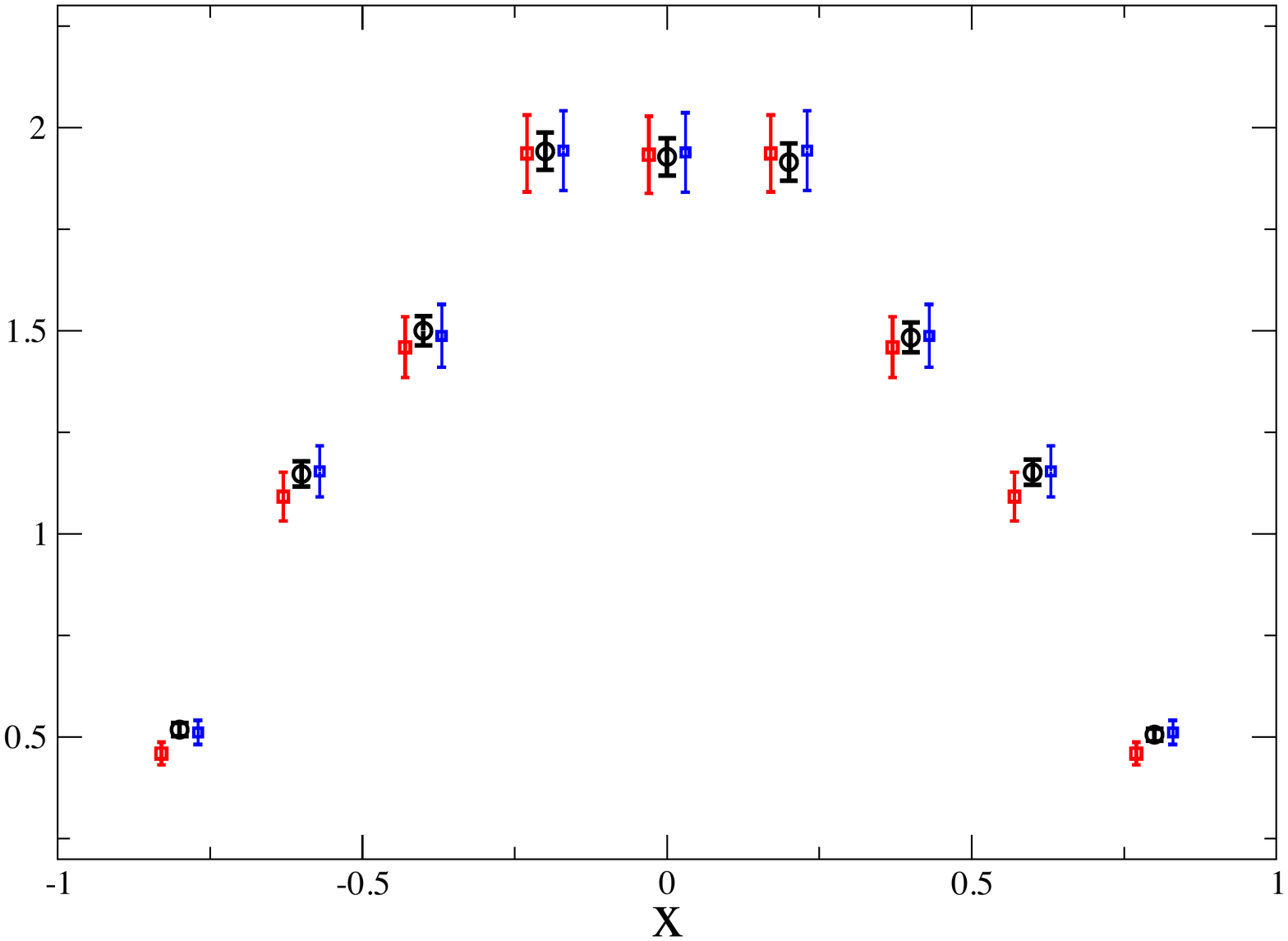}
\includegraphics*[width=0.49\textwidth]{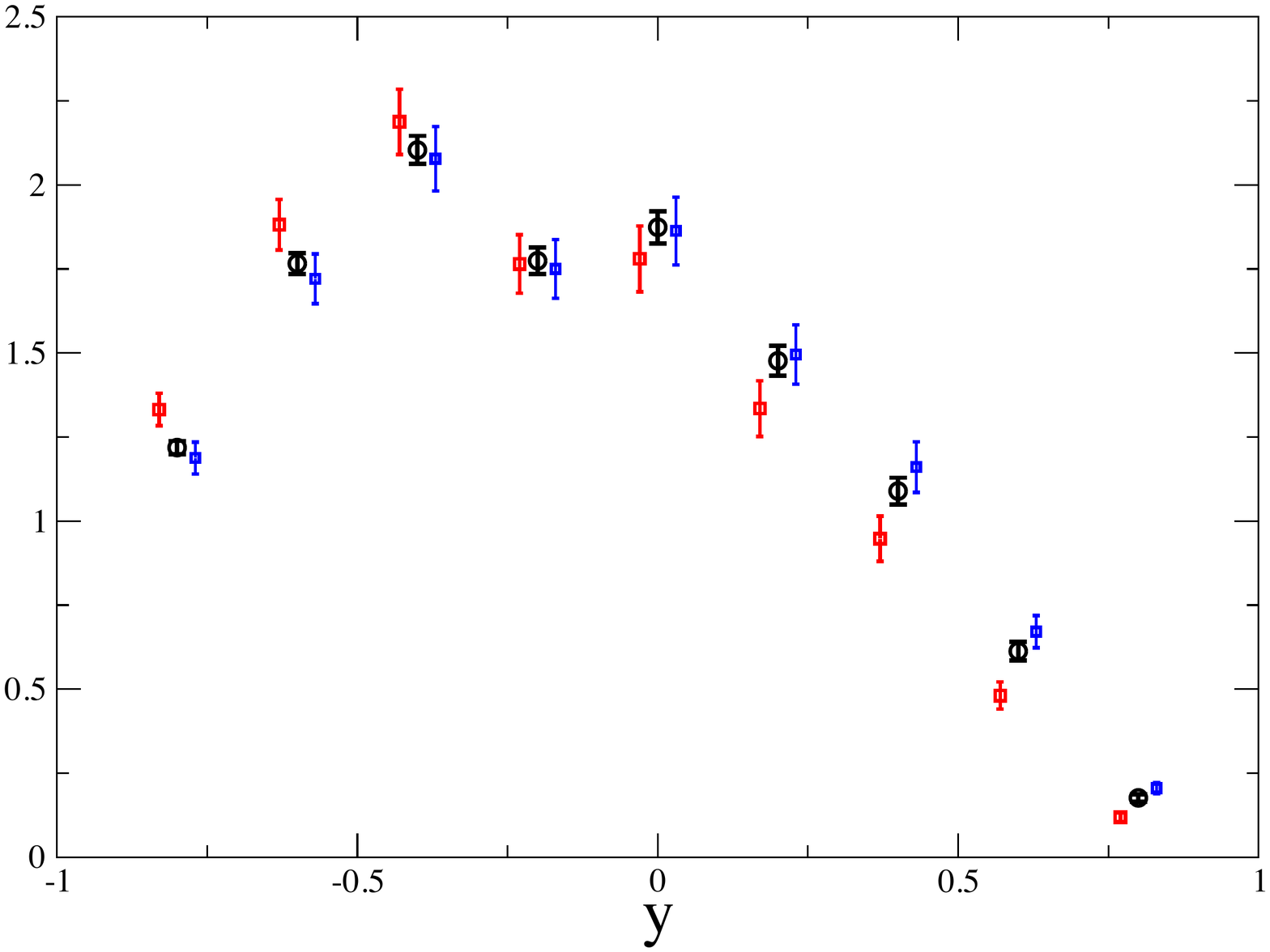}
\includegraphics*[width=0.30\textwidth]{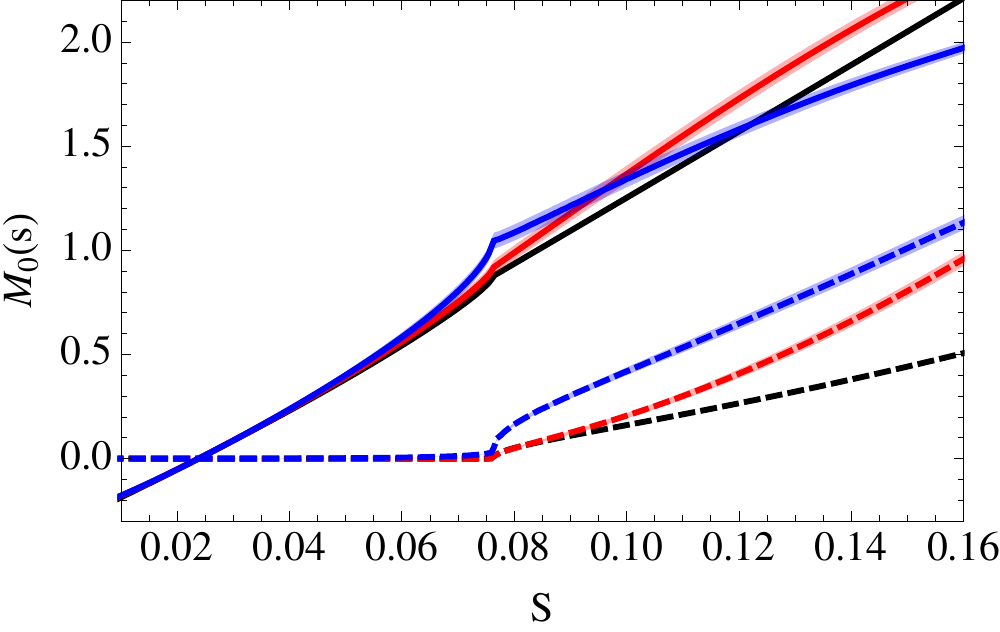}
\includegraphics*[width=0.30\textwidth]{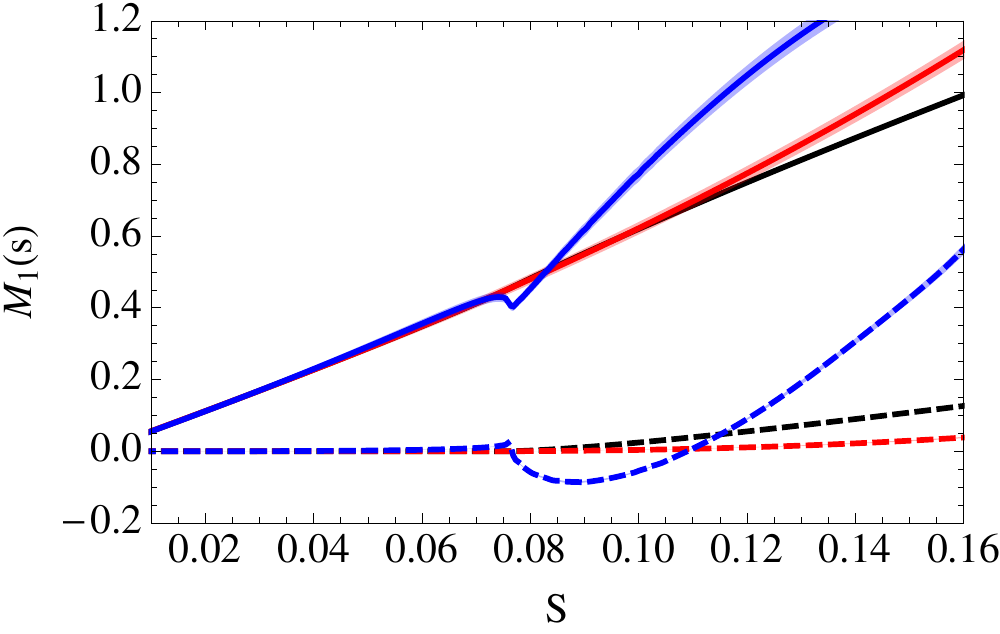}
\includegraphics*[width=0.30\textwidth]{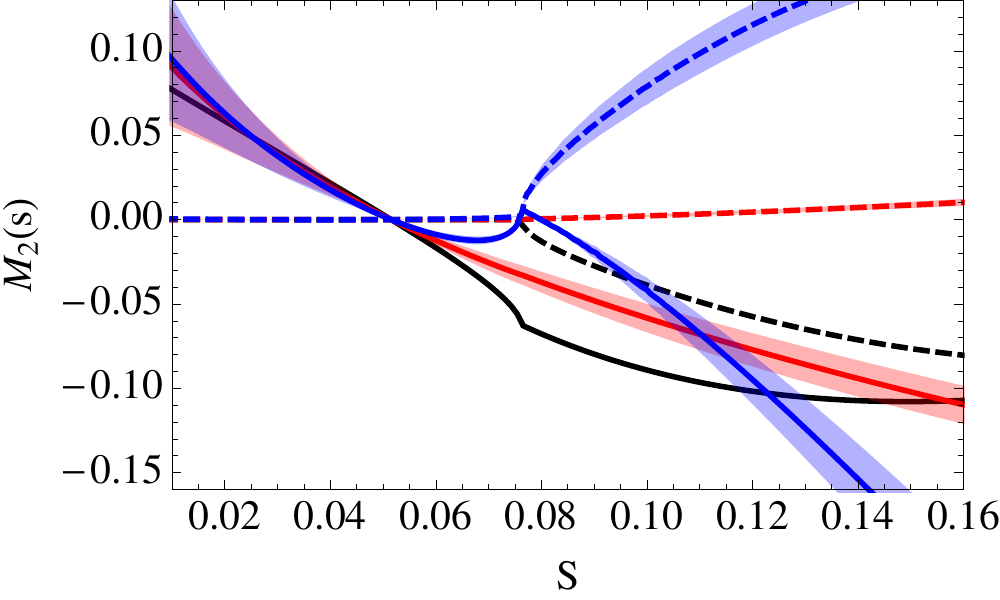}
\caption{Upper panels: $x$- and $y$-projections of the Dalitz plots. Black circles represent the data. The red squares and blue squares are model results using amplitudes with only two-body and including three-body correlations, respectively. The amplitudes were computed using three partial wave components with \mbox{(I,L)=(0,0),\,(2,0),\,(1,1)}. For better visualization fit results are shifted horizontally  (three-body to right and two-body to left) from the experimental points. Bottom panels: The comparison of the NLO $\chi$PT amplitudes $M_{I}$'s  (black curves), with the two-body (red curves) and three-body (blue curves) dispersive amplitudes. Real parts are shown with solid lines and imaginary with dashed lines. In all figures the unknown couplings were fixed by matching to NLO $\chi$PT (see Eq.(\ref{totalMatch})).
\label{fig:Mchipt}}
\end{figure*}

\begin{table}[tbp]
\centering
\caption{Values of $Q$ from different calculations. \label{tab:Q1}}
\renewcommand{\arraystretch}{1.8}
\begin{tabular*}{\columnwidth}{@{\extracolsep{\fill}}llc@{}}
\hline\hline
Theory           & \multicolumn{1}{c}{$Q$}\\
\hline
Set 1 & $21.7 \pm 0.4$  \\
Set 2 & $21.1\pm 0.4$ \\
\hline
Lattice ($N_f=2+1$)\footnote{Here and in the following we combined in quadrature the errors quoted in \cite{Aoki:2013ldr}. } \cite{Aoki:2013ldr}    & $22.6\pm0.9$  \\
\hline
NLO \cite{Gasser:1984pr}    & $20.1$   \\
NNLO \cite{Bijnens:2007pr}     & $22.9$   \\
Kambor \textit{et al.} \cite{Kambor:1995yc}   & $22.4\pm0.9$  \\
Kampf \textit{et al.} \cite{Kampf:2011wr}    & $23.1\pm 0.7$  \\
\hline\hline
\end{tabular*}
\end{table}

As discussed in Sec.~\ref{intro}, the  $\chi$PT  \cite{Bijnens:2007pr} series seems to converge rather slowly and the question arises to which order of  the $\chi$PT should one match the model. It would be desirable to find a matching point where on the $\chi$PT side contributions, from powers of Mandelstam invariants, are small. Therefore, matching the amplitudes in the physical region may not be the best option. Up to NNLO the chiral amplitude satisfies the decomposition of Eq.~(\ref{decayamp}), and up to this order matching is simplified since it is sufficient to match the single variable, partial wave amplitudes \mbox{$a_{IL}(s)$}. The $\chi$PT amplitude for the charged decay, up to NNLO can be written in the form
\begin{align}\label{AchPT}
A^{C}_{\chi PT}(s,t,u) = - \frac{1}{Q^{2}} \frac{m_{K}^{2} (m_{K}^{2} -m_{\pi}^{2})}{3\,\sqrt{3}\,m_{\pi}^{2}\,F_{\pi}^{2}}\,M(s,t,u),
\end{align}
where \mbox{$F_{\pi}=92.3$ MeV}  is the pion decay constant and 
\begin{align}\label{AchPT2}
M(s,t,u) &= M_{0}(s) -\frac{2}{3} M_{2}(s) + M_{2}(t) + M_{2}(u)  \nonumber \\
&+ (s-u) M_{1}(t) + (s-t) M_{1}(u)\,.
\end{align}
Explicit expressions for the functions $M_{I}$ at various orders in the chiral expansion can be found in \cite{Bijnens:2007pr}. Comparing Eq.~(\ref{ampn}) and Eqs. (\ref{AchPT}), (\ref{AchPT2}) one finds
\begin{eqnarray}
a_{00}(s) &=& 3\,N_{\chi PT}\,M_{0}(s),  \nonumber\\ 
a_{20}(s) &=& 2\,N_{\chi PT}\,M_{2}(s),  \\
a_{11}(s) &=& \frac{2}{3}\,N_{\chi PT}\,\frac{K(s)}{s}\,M_{1}(s)\,,\nonumber
\label{Mchipt}
\end{eqnarray}
where 
\begin{equation}
N_{\chi PT}=- \frac{1}{Q^{2}} \frac{m_{K}^{2} (m_{K}^{2} -m_{\pi}^{2})}{3\,\sqrt{3}\,m_{\pi}^{2}\,F_{\pi}^{2}}\,.
\end{equation}

The NNLO $\chi$PT calculation was performed in \cite{Bijnens:2007pr}. The order $O(p^6)$ LECs were estimated using a resonance saturation model and error analysis was not provided. Given that uncertainties in the low energy constants entering $M_I$'s at the NNLOs are not quantitatively settled in the following we choose to match our  dispersive calculation with the NLO $\chi$PT result. In this case one can use the NLO relations between  decay constants and meson masses which reduces the number of low energy constants in the chiral amplitude to one, \mbox{$L_3=(-2.35\pm0.37)\cdot 10^{-3}$} \cite{Amoros:2001cp}. We choose the matching point to coincide with the subtraction point in Eq.~(\ref{gapprox}), which in turn was chosen to coincide with the Adler zero in the LO $\chi$PT amplitude. In that case the determined parameters from matching are the same for the two-body and three-body scenarios.

In the following we  consider two methods for matching the dispersive analysis with $\chi$PT. In the first case we use Eq.~(\ref{Mchipt}) together with the $\chi$PT NLO amplitudes $M_I$'s  to compute the overall normalization and the parameters $g_{IL}(s_0)$, which  in turn  completely determine dispersive amplitudes of our model. We find, 
\begin{align}\label{totalMatch}
g_{00}(s_0)  &= 16.1\,N_{\chi PT}\,,  \nonumber \\
g_{00}(s_0)/ g_{20}(s_0)/g_{11}(s_0)  &= 1/0.12/(0.129\pm0.014)\,.
\end{align} 
This confirms that the amplitude \mbox{$(I,L)=(0,0)$} is dominant. In the lower panel of Fig.~\ref{fig:Mchipt}, we compare the $\chi$PT amplitudes with the dispersive ones, the latter obtained using the subtraction constants from Eq.~(\ref{totalMatch}). Comparing with the WASA-at-COSY data shown in the upper panel in Fig.~\ref{fig:Mchipt}, we find that the dispersive amplitude fixed by Eq.~(\ref{totalMatch}) gives \mbox{$\chi^{2}/d.o.f.$} of approximately $13.0$ using only two-body amplitudes, which is reduced to $2.9$  when three-body rescattering contributions are included. Even though the model compares reasonably well with the data, the large value of \mbox{$\chi^{2}/d.o.f.$} prevents us from extracting the $Q$-value using this method.

To extract the $Q$-value we therefore use the $\chi$PT amplitudes to determine the overall normalization only,  while for the subtraction constants $g_{IL}(s_0)$ we use the results from the fit of the  WASA-at-COSY data described in the previous section. We find  \mbox{$Q(\text{Set 1})=21.7 \pm 0.4$} and \mbox{$Q(\text{Set 2})=21.1\pm 0.4$} for the two sets of parameters given in Table ~\ref{tab:xy}. Comparison of our findings with previous results is summarized in Table \ref{tab:Q1}. We observe that the extracted $Q$-values are somewhat smaller compared to \cite{Bijnens:2007pr,Kambor:1995yc,Kampf:2011wr}, and within $1\sigma$ from the recent ($N_f=2+1$) lattice computations \cite{Aoki:2013ldr}. We note that lattice calculations of electromagnetic correction for  $N_f=2+1$ are not yet available, while for $N_f=2$  these were reported in \cite{Divitiis:2013ct}. The lattice result given in Table \ref{tab:Q1} depends on the input value for the light quark mass ratio,  $m_u/m_d=0.46\pm0.03$ which is the LO $\chi$PT result reduced by a factor of 8(4)\%  chosen as an estimate of the correction from higher-orders chiral effects \cite{Aoki:2013ldr}. Alternatively, using the extracted $Q$-value and the  $N_f=2+1$  lattice result for $m_s/\hat{m}=27.46\pm0.44$ \cite{Aoki:2013ldr} we can estimate $m_u/m_d$. We find 
\begin{equation}
\frac{m_u}{m_d}=0.42\pm0.02
\end{equation}
as an average between Sets 1 and 2. Another useful quantity that can be calculated from our  $Q$ and $m_s/\hat{m}$ is the so-called $R$-value given by
\begin{eqnarray}
R=\frac{m_s-\hat{m}}{m_d-m_u}&=&2\,Q^2\left(1+\frac{m_s}{\hat{m}}\right)^{-1}\nonumber\\
&=&32.2\pm 1.3\,.
\end{eqnarray}

\section{Conclusions}
\label{SectionIV}

In this paper, a new data driven dispersive analysis of $\eta\rightarrow 3\pi$ was performed. The hadronic final state interactions were incorporated using the  Khuri-Treiman equation, which was solved using Pasquier inversion technique. To the best of our knowledge it is the first time such an approach has been used in analysis of the $\eta$ decays. In an earlier study \cite{Guo:2014vya}, we illustrated the pros and cons of the Pasquier technique using a toy model with known exact solutions. The main limitation of this method is related to the treatment of the left-hand cuts, which in general are not known. We approximated them by a constant which is absorbed in the subtraction constants. As it was shown in \cite{Guo:2014vya}, this approximation works very well, when the physical region does not depend strongly on the accurate form of the left-hand cut.
On the other hand the advantage of the Pasquier inversion  is that it eliminates the  need for specifying the  high-energy behavior of the absorptive parts in the physical region.

In the analysis of the $\eta\to3\pi$ decays presented here, we have shown that with a single real parameter ($g_{11}$) and the physical $\pi\pi$ partial-wave amplitudes \cite{GarciaMartin:2011cn} it is possible to reproduce the Dalitz distribution of the charged $\eta$ decay mode \cite{Adlarson:2014aks}.  We have also verified that including more partial waves leads fits with comparable $\chi^2/d.o.f.$  The resulting Dalitz parameters, averaged over the various combinations of partial waves considered in this paper are, 
\begin{eqnarray}
&&a=1.116\pm0.032\,,\quad b=0.188\pm0.012\,,\nonumber \\
&&d=0.063\pm0.004\,,\quad f=0.091\pm0.003\,,\\
&&g=0.042\pm0.009\,.\nonumber
\end{eqnarray}
These are consistent, within $1\sigma$ with the analysis of WASA-at-COSY having central values shifted towards values obtained from analysis by the KLOE Collaboration, which were not include in our fits. Based on the analysis of the charged decay we made a prediction for the  slope parameter of the Dalitz distribution in the neutral decay channel, 
\begin{equation}
\alpha=-0.022\pm0.004\,.
\end{equation}
This value is above the PDG value of $\alpha^{exp}=-0.0317\pm0.0016$. We speculate that the discrepancy may be a consequence of the WASA-at-COSY $b$ parameter being significantly larger than in the earlier KLOE analysis \cite{Ambrosino:2008ht}. We expect that in the future this issue will be resolved once the new KLOE data \cite{Balkestahl:2015aka} become available allowing a simultaneous fit of both data sets.

Another useful test of the amplitudes is  provided by the ratio of neutral and charged decay rates. In the isospin limit this ratio does not depend on the normalization, and if the small electromagnetic isospin breaking is also ignored~\cite{Ditsche:2008cq}, it depends only on the integrated Dalitz plot distributions. From our amplitude we find 
\begin{equation}
r=\frac{\Gamma(\eta\rightarrow 3\pi^0)}{\Gamma(\eta\rightarrow \pi^+\pi^-\pi^0)}=1.52 \pm 0.09\,,
\end{equation}
which is consistent with the experimental value of $r^{exp}=1.43\pm0.02$ \cite{PDG-2012}. We have also compared our amplitudes with the NLO  $\chi$PT results and found the $Q$-value of 
\begin{equation}
Q=21.4\pm 0.4\,.
\end{equation}
% consistent with the recent lattice results. 
The error is of the statistical origin. It was computed through standard error propagation of the uncertainties arising from the $\pi\pi$ phase shifts, the $L_3$ coefficient, the experimental decay width $\Gamma(\eta\rightarrow \pi^+\pi^-\pi^0)$ and the statistical error in fitting the Dalitz plot. Inelasticity and higher partial waves are also potential sources of uncertainties \cite{Guo:2015kla}.

Using the extracted $Q$-value and recent averages from the $N_f=2+1$ lattice computation for $\hat{m}=3.42\pm0.09$ and $m_s=93.8\pm0.24$,  \cite{Aoki:2013ldr}  we estimate 
the  up and down quark masses to be
\begin{eqnarray}
m_u&=&2.02\pm0.14~\text{MeV}\,, \nonumber\\
m_d&=&4.82\pm0.08~\text{MeV}.
\end{eqnarray}

The method for amplitude construction presented in this work can be directly applied to decays of  heavier meson, {\it e.g.} $\eta'$ and used, for example,  to test reliability of the isobar model. It can also be extended to incorporate couple-channels, which might be more relevant in decays of heavier mesons.

All the material, including data and code are available in an interactive form online~\cite{website}.

\begin{acknowledgments}
We would like to thank B.~Kubis, V.~Mokeev, E.~Passemar and M.~R.~Pennington for useful discussions. In addition I.~V.~D. acknowledges discussions with G.~Colangelo and H.~Leutwyler. This material is based upon work supported in part by the U.S. Department of Energy, Office of Science, Office of Nuclear Physics under contract DE-AC05-06OR23177. This work was also supported in part by the U.S. Department of Energy under Grant No. DE-FG0287ER40365, National Science Foundation under Grants PHY-1415459 and PHY-1205019, and IU Collaborative Research Grant.

After submission of our manuscript a new $\eta\rightarrow 3\pi$ analysis by the BESIII Collaboration became available \cite{Ablikim:2015cmz}. The values for the Dalitz plot parameters (except the parameter $f$) of the charged $\eta$ decay are compatible with our results within the error bars. We note that the determined $b^{\text{BESIII}}=0.153\pm0.017\pm0.004$ is considerably lower than WASA-at-COSY result. It confirms the expected correlation to the slope parameter in the neutral decay channel, which turned out to be $\alpha^{\text{BESIII}}=-0.055\pm0.014\pm0.004$.

\end{acknowledgments}

\vspace*{1cm}

\appendix
\section{Isospin algebra}
\label{App:A}

In Eq.~(\ref{decayamp}) the isospin factors are given by 
\begin{align}
& \mathcal{P}^{(0)}_{\alpha \beta \gamma \eta} = \frac{1}{3}\,\delta_{\alpha \beta} \,\delta_{\gamma \eta} \,, \nonumber \\
& \mathcal{P}^{(1)}_{\alpha \beta \gamma \eta} = \frac{1}{2}\, (  \delta_{\alpha \gamma} \,\delta_{  \beta \eta} - \delta_{\alpha \eta }\,\delta_{ \beta \gamma }  ) \,,  \\
& \mathcal{P}^{(2)}_{\alpha \beta \gamma \eta} = \frac{1}{2} \,(  \delta_{\alpha \gamma} \,\delta_{  \beta \eta} + \delta_{\alpha \eta }\, \delta_{ \beta \gamma }  )- \frac{1}{3} \, \delta_{\alpha \beta}\, \delta_{\gamma \eta}\,, \nonumber
\end{align}
which satisfy, 
\begin{eqnarray}
&&\sum_{\eta\gamma} \mathcal{P}^{(I)}_{\alpha \beta \eta\gamma}\mathcal{P}^{(I')}_{ \eta\gamma\alpha' \beta'}=\mathcal{P}^{(I')}_{\alpha \beta\alpha'\beta'}\delta_{II'}\,,\nonumber\\
&&\sum_{\eta\gamma} \mathcal{P}^{(I)}_{\beta\gamma\alpha\eta}\mathcal{P}^{(I')}_{ \eta\gamma\alpha' \beta'}=\mathcal{P}^{(I')}_{\alpha \beta\alpha'\beta'}[C_{st}]_{II'}\,,\\
&&\sum_{\eta\gamma} \mathcal{P}^{(I)}_{\gamma\alpha\beta\eta}\mathcal{P}^{(I')}_{ \eta\gamma\alpha' \beta'}=\mathcal{P}^{(I')}_{\alpha \beta\alpha'\beta'}[C_{su}]_{II'}\,.\nonumber
\end{eqnarray}
Here $\alpha,\,\beta,\,\gamma,\,\eta$ are the Cartesian isovector indices and isospin crossing matrices $C_{st}$ and $C_{su}$, are given by 
\begin{eqnarray}
&&\hspace{-1em}
C_{st}=
\left(\!\!\!\begin{array}{ccc}
1/3&1&5/3\\\rule{0em}{1em}
1/3&1/2&-5/6\\
\rule{0em}{1em}
1/3&-1/2&1/6\end{array}
\!\right),\quad
C_{su}=
\left(\!\!\!\!\begin{array}{ccc}
1/3&-1&5/3\\\rule{0em}{1em}
-1/3&1/2&5/6\\
\rule{0em}{1em}
1/3&1/2&1/6\end{array}
\!\right) \,.
\nonumber
\end{eqnarray}

\section{Kernel functions}
\label{App:B}

\begin{figure}[tbp]
\includegraphics[width=0.48\textwidth]{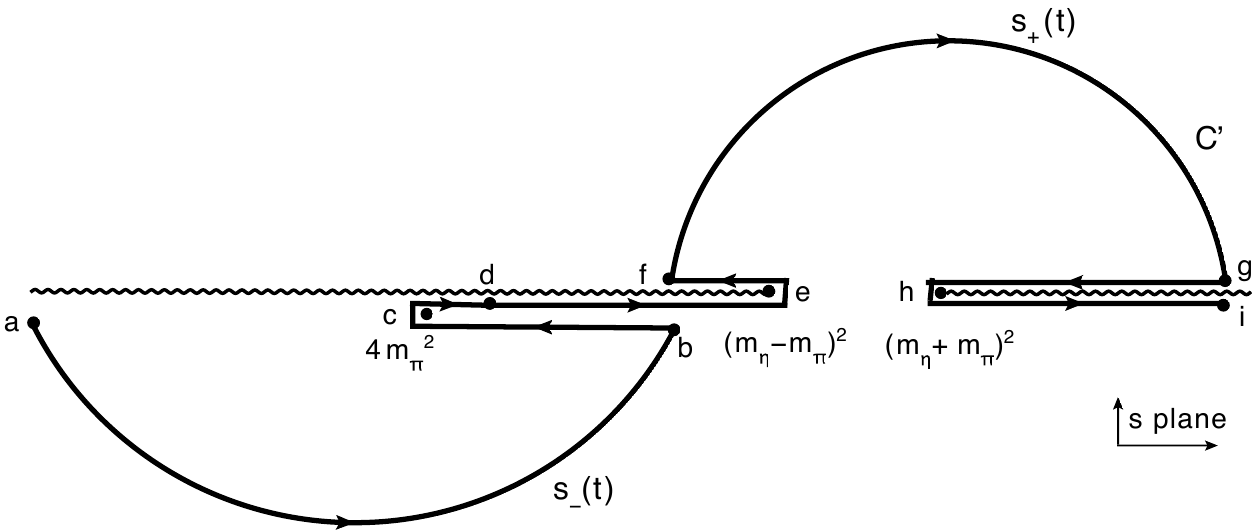}
\caption{Integration contour $C'$ in the complex $s$ plane after Pasquier inversion. The  black wiggle lines represent cuts attached to two branch points: $(m_{\eta}\pm m_{\pi})^{2}$ in $s$-plane. The  points labeled by \mbox{$a-i$} correspond to (a) \mbox{$s_{-}(0)= -\infty$}, (b)\mbox{$s_{-}(4 m_{\pi}^{2})= \frac{ m_{\eta}^{2} -m_{\pi}^{2} }{2}$},  (c) \mbox{$s_{-}(\frac{ m_{\eta}^{2} -m_{\pi}^{2} }{2})= 4 m_{\pi}^{2}  $}, (d) \mbox{$s_{\pm}((m_{\eta}-m_{\pi})^{2})= m_{\pi} ( m_{\pi}+ m_{\eta}) $}, (e) \mbox{$s_{+}(m_{\pi}(m_{\eta}+m_{\pi}))= (m_{\eta}-m_{\pi})^{2} $}, (f) \mbox{$s_{+}(4m_{\pi}^{2})= \frac{ m_{\eta}^{2} -m_{\pi}^{2} }{2}$},  (g) \mbox{$s_{+}(0)= \infty$}, (h) \mbox{$s_{+}(m_{\pi}(m_{\pi}-m_{\eta}))= (m_{\eta}+m_{\pi})^{2}$},  and (i) $s_{+}(-\infty)= \infty$, respectively.\label{fig:Contour_2}}
\end{figure} 

The kernel functions in Eq. (\ref{pasqg}) are determined as
\begin{align}
&\mathcal{K}_{IL, I'L'} (s,t) = 2\,(2 L'+1)   \nonumber \\
& \quad \quad  \times  \left(    \theta(t)\,\Delta_{I L, I' L'}(s,t)-    \theta(-t)\,\Sigma_{IL, I'L'}(s,t) \right) ,
\end{align}
with  
 \begin{align}
&\Delta_{IL, I'L'}(s,t)  =   \int_{s_{-}(t)}^{s_{+}(t)} \!\!\!\!\!\!\!\!\!\!\!\! (C') \  \frac{d s'}{s'-s}  \frac{ \rho(s')\,(s'/4- m_{\pi}^{2})^{L}}{ \mathcal{F}_{IL}(s')\,K^{L+1}(s')/s'}      \nonumber \\
& \quad  \quad \quad \quad \quad \quad     \times  \frac{ \mathcal{F}_{I'L'}(t)\,K^{L'}(t)}{(t/4- m_{\pi}^{2})^{L'}} P_{L}(z_{s'})\,P_{L'}(z_{t})   ,  
\end{align}
and
\begin{align}
&\Sigma_{IL,I'L'}(s,t)  =    \int_{s_{+}(t)}^{ \infty} \!\!\!\!\!\!\!\!\!\! (C') \  \frac{d s'}{s'-s}  \frac{\rho(s')\,(s'/4- m_{\pi}^{2})^{L}}{ \mathcal{F}_{IL}(s')\,K^{L+1}(s')/s'}      \nonumber \\
& \quad  \quad \quad \quad \quad \quad     \times  \frac{  \mathcal{F}_{I'L'}(t)\,K^{L'}(t)}{(t/4- m_{\pi}^{2})^{L'}} P_{L}(z_{s'})\,P_{L'}(z_{t})   ,
\end{align}
where the contour $C'$ is shown in Fig.~\ref{fig:Contour_2} (see \cite{Guo:2014vya} for more details). These kernel functions can be computed analytically, what significantly speeds up numerical computations. In the calculations presented in this paper, only the functions  $\Delta_{IL, I'L'}(s,t)$ are needed and their analytical representations are below in terms of
\begin{align}
&\Delta_{L, L'}(s,t)  \equiv  K^{L'}(t)   \int_{s_{-}(t)}^{s_{+}(t)} \!\!\!\!\!\!\!\!\!\!\!\! (C') \  \frac{d s'}{s'-s}  \frac{1}{ U(s')}      \nonumber \\
& \quad  \quad \quad \quad \quad       \times  \frac{(s'- 4\,m_{\pi}^{2})^{L}}{K^{L}(s')} P_{L}(z_{s'})\,P_{L'}(z_{t})\,,  
\end{align}
so that for \mbox{$L=0$},
\begin{align}
&\Delta_{I0, I'L'}(s,t)    = 4^{L'-L}\frac{ \mathcal{F}_{I'L'}(t) }{ (t- 4\,m_{\pi}^{2})^{L'}}  \\
& \quad \quad \times \left [ \frac{\Delta_{0, L'}(s,t)}{\mathcal{F}_{I 0} (s)}  -  \frac{s_{\chi}^{(I)} - s_{A}^{(I)}}{s-s_{\chi}^{(I)}} \Delta_{0,L'} ( s_{\chi}^{(I)},t)  \right ]\,,\nonumber
\end{align}
and otherwise \mbox{$(L\neq 0)$}, 
\begin{align}
&\Delta_{IL, I'L'}(s,t)    = 4^{L'-L} \frac{ \mathcal{F}_{I'L'}(t) }{ (t- 4\,m_{\pi}^{2})^{L'}}\,\Delta_{L, L'}(s,t)\,.  
\end{align}
The square root function $U(z)$ is given by  
\begin{equation}
U(z)=\sqrt{(z - (m_{\eta}-m_{\pi})^{2} )(z -(m_{\eta}+m_{\pi})^{2})}
\end{equation}
in the complex $z$ plane. Here and in what follows, the phase convention for $U(z)$ is chosen by \mbox{$U(s\pm  i0 ) = (\mp , i , \pm)\,|U(s)|$} for \mbox{$s\in ((-\infty, (m_{\eta}-m_{\pi})^{2} ]$}, \mbox{$[ (m_{\eta}-m_{\pi})^{2} , (m_{\eta}+m_{\pi})^{2} ], [ (m_{\eta}+m_{\pi})^{2} , \infty))$} respectively.  The kinematic factor \mbox{$K(s)/(s\,\rho(s))$} is given by   the value of $U(s)$   right below the two cuts attached to branch points $s=(m_{\eta}\pm m_{\pi})^{2}$, {\it i.e.} \mbox{$K(s)/(s\,\rho(s)) =U(s-i0) $}. For real $s$ and $t$ the physical values of \mbox{$\Delta_{L,L'} (s,t)$}   correspond to the limit  \mbox{$s + i0$} and \mbox{$t + i0$}. 

%The kernel functions $\Delta_{L, L'}$ are given by:
\begin{itemize} \item $\bf{(L,L')=(0, 0)}$:
\begin{align}
\Delta_{0,0} ( s,t) &  = \frac{1}{U(s )}  \left[ \ln  \left | \frac{R  (s, t) + U(s)\,U(t)}{R  (s, t) - U(s)\,U(t) } \right |   -  i\,\pi\,\theta \left ( \varphi  \left (s, t \right ) \right )  \right] ,\nonumber  \\
R (s, t)  &= - m^{4}_{\eta}  +  (s - m_{\pi}^{2}) (t- m_{\pi}^{2})    +m_{\eta}^{2}\,(s+t), \nonumber \\
\varphi(s,t)&=s\,t\, (m_{\eta}^{2} + 3 m_{\pi}^{2} -s -t)- m_{\pi}^{2} \,(m_\eta^{2}-m_{\pi}^{2})^{2}.\nonumber 
\end{align}
\\

\item $\bf{(L,L')=(0, 1)}$:
\begin{align}
&  \Delta_{0,1} ( s,t)    = 2\,t\,\Delta_{a} ( t) +t\,(2 s + t - m_{\eta}^{2} - 3 m_{\pi}^{2})\, \Delta_{0,0} ( s,t) ,\nonumber  \\
& \Delta_{a} ( t) = \int_{s_{-}(t)}^{s_{+}(t)}  \!\!\!\!\!\!\!\!\ \!\!\!\!\!\!  (C') \  \frac{d s'}{  U(s')  }  =  -   \ln   \frac{ \left (  m_{\eta}^{2} - m_{\pi}^{2} + t + U(t) \right )^{2} }{4\,m_{\eta}^{2}\,t}. \nonumber
\end{align}

\item $\bf{(L,L')=(1, 0)}$:

\begin{align}
&\Delta_{1,0}(s,t)    = \frac{1}{s} \left (\overline{\Delta}_{1,0} (s, t)  -\overline{\Delta}_{1,0} (0, t)  \right), \nonumber\\
& \overline{\Delta}_{1,0} (s, t)    =  \Delta_{0,0} ( s,t) + 2\,(t+ m_{\pi}(m_{\eta} - m_{\pi}) )\,\Delta^{(+)}_{b} (s, t)   \nonumber \\
&\quad\quad \quad\quad + (m_{\eta} -m_{\pi})^{2}  \Delta^{(-)}_{b} (s, t)  \nonumber \\
&\quad\quad\quad\quad  + 2\,(t+ m_{\pi}(m_{\eta} - m_{\pi}) ) (m_{\eta} -m_{\pi})^{2}   \Delta_{c} ( s,t) , \nonumber\\
& \Delta^{(\pm)}_{b} (s,t) = \int_{s_{-}(t)}^{s_{+}(t)}  \!\!\!\!\!\!\!\!\ \!\!\!\!\!\!  (C') \  \frac{d s'}{s'-s}  \frac{ 1 }{  U(s')  }  \frac{1}{s' -(m_{\eta} \pm m_{\pi})^{2} }  \nonumber \\
&\quad\quad  \quad\quad  =  \frac{1}{s -(m_{\eta} \pm m_{\pi})^{2}  }   \left( \Delta_{0,0} ( s,t)   -  \Delta_{d}^{(\pm)} (  t)    \right) , \nonumber\\
& \Delta_{c} (s,t) = \int_{s_{-}(t)}^{s_{+}(t)}  \!\!\!\!\!\!\!\!\ \!\!\!\!\!\!  (C') \   \frac{d s' }{s'-s}  \frac{ 1 }{  U^{3}(s')  }   \nonumber \\
&\quad\quad \quad  =  \frac{\Delta_{0,0} ( s,t) }{ U^{2}(s)  }   + \frac{1}{4\,m_{\pi}\,m_{\eta} } \frac{ \Delta_{d}^{(-)} (t)  }{s -(m_{\eta} - m_{\pi})^{2}  }   \nonumber \\
&\quad\quad\quad  - \frac{1}{4\,m_{\pi}\,m_{\eta} } \frac{ \Delta_{d}^{(+)}(t)  }{s -(m_{\eta} + m_{\pi})^{2}  } , \nonumber\\
&  \Delta_{d}^{(\pm)}   (t) = \int_{s_{-}(t)}^{s_{+}(t)}  \!\!\!\!\!\!\!\!\ \!\!\!\!\!\!  (C') \   \frac{d s'}{s' -(m_{\eta} \pm m_{\pi})^{2} }   \frac{ 1 }{  U(s')  }    \nonumber \\
&  \quad \quad \quad = \frac{U(t)}{m_{\eta}\,(m_{\eta} \pm m_{\pi}) (t \pm m_{\pi} (m_{\eta} \mp m_{\pi}))}\,.\nonumber
\end{align}

\item $\bf{(L,L')=(1, 1)}$:
\begin{align}
&\Delta_{1,1} ( s,t)  = \frac{t}{s} \left(\overline{\Delta}_{1,1} (s, t)  -\overline{\Delta}_{1,1} (0, t)  \right), \nonumber\\
& \overline{\Delta}_{1,1} (s, t)    =  2\,\Delta_{a} ( t)   + 4\,( t+ m_{\pi}(m_{\eta} - m_{\pi}) )  \Delta_{d}^{(+)}(t) \nonumber \\
&\quad\quad \quad\quad +2\,(m_{\eta} -m_{\pi})^{2}  \Delta_{d}^{(-)}   (   t)   \nonumber \\
&\quad\quad \quad\quad  +4\,( t+ m_{\pi}(m_{\eta} - m_{\pi}) ) (m_{\eta} -m_{\pi})^{2}   \Delta_{e} ( t)  \nonumber \\
&\quad\quad \quad\quad + (2 s + t-m_{\eta}^{2} -3 m_{\pi}^{2})\,\overline{\Delta}_{1,0} (s, t),\nonumber\\
 & \Delta_{e} ( t) = \int_{s_{-}(t)}^{s_{+}(t)}  \!\!\!\!\!\!\!\!\ \!\!\!\!\!\!  (C') \   \frac{d s' }{  U^{3}(s')  }     =  \frac{\Delta_{d}^{(+)} ( t)  -  \Delta_{d}^{(-)} ( t)   }{4\,m_{\pi}\,m_{\eta} }  .\nonumber
\end{align}

\item $\bf{(L,L')=(0, 2)}$:
 \begin{align}
& \Delta_{0,2} ( s,t)    =  6\,t^{2}\,\Delta_{f} ( s,t)   + 6\,t^{2}\,(2s+t-m^{2}_{\eta} -3 m^{2}_{\pi} )\, \Delta_{a}(t)      \nonumber \\
&\quad\quad \quad\quad + \frac{ 3\,t^{2}\,(2\,s+t-m^{2}_{\eta} -3\, m^{2}_{\pi} )^{2}  -U^{2}(t)}{2}  \Delta_{0,0} ( s,t) ,\nonumber\\
&  \Delta_{f} (s, t) = \int_{s_{-}(t)}^{s_{+}(t)}  \!\!\!\!\!\!\!\!\ \!\!\!\!\!\!  (C') \  \frac{ d s' }{  U(s')  }( s'-s )    \nonumber\\
&\quad\quad \quad\quad =  U(t) + (m_{\eta}^{2} + m_{\pi}^{2} -s)\,\Delta_{a} ( t). \nonumber
\end{align}

\item $\bf{(L,L')=(2, 0)}$:
\begin{align}
&\Delta_{2,0} ( s,t)   = -\frac{1}{2}\,\Delta_{i}  ( s,t) + \frac{3}{2}\, \Delta_{j}^{(-)}  (s, t)  \nonumber \\
 & \quad\quad \quad\quad + 6\,(t + m_{\pi}\,(m_{\eta} -m_{\pi})) \frac{  \Delta_{j}^{(-)}  (s, t)  -    \Delta_{l}^{(-)}  ( t) }{s- (m_{\eta}+ m_{\pi})^{2}} \nonumber \\
 & \quad\quad \quad\quad + \frac{3\,(t + m_{\pi} (m_{\eta} -m_{\pi}))^{2} }{2\,m_{\eta} m_{\pi}} \frac{  \Delta_{j}^{(+)}  (s, t)  -    \Delta_{l}^{(+)}  ( t) }{s- (m_{\eta}- m_{\pi})^{2}} \nonumber \\
 & \quad\quad \quad\quad - \frac{3\,(t + m_{\pi} (m_{\eta} -m_{\pi}))^{2} }{2\,m_{\eta} m_{\pi}} \frac{  \Delta_{j}^{(-)}  (s, t)  -    \Delta_{l}^{(-)}  (t) }{s- (m_{\eta}+ m_{\pi})^{2}},\nonumber\\
&\Delta_{i}  ( s,t)= \int_{s_{-}(t)}^{s_{+}(t)}  \!\!\!\!\!\!\!\!\ \!\!\!\!\!\!  (C') \  \frac{d s' }{s'(s' -s) }   \frac{ s'-4\,m_{\pi}^{2} }{  U^{3}(s')  }      \nonumber \\
&\quad\quad \quad  =  \left(1-\frac{4\,m_{\pi}^{2}}{s}\right) \Delta_{c}  (s, t) +\frac{4\,m_{\pi}^{2}}{s}  \Delta_{c}(0, t),\nonumber\\
& \Delta_{j}^{(\pm)}  (s, t) = \int_{s_{-}(t)}^{s_{+}(t)}  \!\!\!\!\!\!\!\!\ \!\!\!\!\!\!  (C') \   \frac{d s'}{s' -s }   \frac{ 1 }{  U(s')  }  \frac{1}{ (s'- (m_{\eta} \pm m_{\pi})^{2})^{2}}     \nonumber \\
  &\quad\quad \quad\quad =   \frac{  \Delta_{0,0}  ( s,t)  -  \Delta_{d}^{(\pm)}  ( t)}{ (s- (m_{\eta} \pm m_{\pi})^{2})^{2}}     -   \frac{    \Delta_{k}^{(\pm)}  ( t)}{ s- (m_{\eta} \pm m_{\pi})^{2}},\nonumber\\
&\Delta_{k}^{(\pm)}  ( t) = \int_{s_{-}(t)}^{s_{+}(t)}  \!\!\!\!\!\!\!\!\ \!\!\!\!\!\!  (C') \   \frac{d s' }{  U(s')  }   \frac{1}{ (s'- (m_{\eta} \pm m_{\pi})^{2})^{2}}     \nonumber \\
  &       =   \pm \frac{4}{3} \frac{m_{\pi} (m_{\eta} \pm m_{\pi}) (t \pm m_{\pi} (m_{\eta} \mp m_{\pi}))}{(4 m_{\eta} m_{\pi})^{2}}   \nonumber \\
  &     \times \frac{U(t)}{ \varphi( (m_{\eta} \pm m_{\pi})^{2},t)} \left( 3 +\frac{m_{\pi}^{2}U^{2}(t) + 3 m_{\eta}^{2} t (t-4 m_{\pi}^{2})}{ \varphi( (m_{\eta} \pm m_{\pi})^{2},t)}   \right) ,\nonumber\\
 & \Delta_{l}^{(\pm)}  (t) = \int_{s_{-}(t)}^{s_{+}(t)}  \!\!\!\!\!\!\!\!\ \!\!\!\!\!\!  (C') \    \frac{d s' }{  U^{3}(s')  }  \frac{1}{ s'- (m_{\eta} \pm m_{\pi})^{2}}     \nonumber \\
  &\quad\quad \quad  =        \pm    \frac{    \Delta_{k}^{(\pm)}   ( t) -  \Delta_{c}(t)}{ 4 m_{\eta} m_{\pi}}.\nonumber
\end{align}

\item $\bf{(L,L')=(2, 1)}$:

\begin{align}
&\Delta_{2,1}(s,t) = t\,(2\,s + t - m_{\eta}^{2} - 3\,m_{\pi}^{2})  \Delta_{2,0}  ( s,t)  \nonumber \\
&\quad\quad \quad\quad  -t\,\Delta_{e}(t) + 4\,m_{\pi}^{2}\,t\,\Delta_{c} ( 0,t)   \nonumber \\
& \quad\quad \quad\quad + 3\,t\,\Delta_{k}^{(-)}  (t)  + 6\,t\,(t + m_{\pi} (m_{\eta} - m_{\pi}))\,\Delta_{l}^{(-)}  (t)  \nonumber \\
& \quad\quad \quad\quad + 6\,t\,(t + m_{\pi} (m_{\eta} - m_{\pi}))^{2}  \Delta_{m} (t) ,\nonumber\\
&  \Delta_{m} (t) = \int_{s_{-}(t)}^{s_{+}(t)}  \!\!\!\!\!\!\!\!\ \!\!\!\!\!\!  (C') \    \frac{d s' }{  U^{5}(s')  }    =     \frac{  \Delta_{l}^{(+)}  ( t)  -  \Delta_{l}^{(-)}  ( t)}{ 4\,m_{\eta}\,m_{\pi}   }.\nonumber
\end{align}

\item $\bf{(L,L')=(1, 2)}$:
 \begin{align}
& \Delta_{1,2} ( s,t)    = \frac{1}{s} \left(\overline{\Delta}_{1,2} (s, t)  -\overline{\Delta}_{1,2} (0, t)  \right), \nonumber\\
&\overline{\Delta}_{1,2} ( s,t)   =  \bigg [1+ \frac{2\,(t + m_{\pi}(m_{\eta}-m_{\pi}))}{s-(m_{\eta}+m_{\pi})^{2}} + \frac{  (m_{\eta}-m_{\pi})^{2}}{s-(m_{\eta}-m_{\pi})^{2}}   \nonumber \\
& \quad\quad + \frac{2\,(t + m_{\pi}(m_{\eta}-m_{\pi}))  (m_{\eta}-m_{\pi})^{2} }{U^{2}(s)} \bigg ] \Delta_{0,2} ( s,t)  \nonumber \\
&\quad\quad - \frac{1}{4 m_{\pi } m_{\eta}} \frac{2\,(t + m_{\pi}(m_{\eta}-m_{\pi}))  (m_{\eta}+m_{\pi})^{2} }{s-(m_{\eta}+m_{\pi})^{2}}   \Delta_{g}^{(+)} ( t)  \nonumber \\
&\quad\quad + \frac{1}{4 m_{\pi } m_{\eta}} \frac{2\,(t - m_{\pi}(m_{\eta}+m_{\pi}))  (m_{\eta}-m_{\pi})^{2} }{s-(m_{\eta}-m_{\pi})^{2}}   \Delta_{g}^{(-)} ( t)  ,\nonumber \\
&\Delta_{g}^{(\pm)}(t) = \int_{s_{-}(t)}^{s_{+}(t)}  \!\!\!\!\!\!\!\!\ \!\!\!\!\!\!  (C') \  \frac{d s' }{s' -(m_{\eta} \pm m_{\pi})^{2} }   \frac{ 1 }{  U(s')  }    \nonumber \\
  & \quad\quad \quad\quad  \times \frac{ 3\,t^{2} (2s'+t-m^{2}_{\eta} -3\,m^{2}_{\pi} )^{2}  -U^{2}(t)}{2}    \nonumber \\
  &\quad\quad  \quad  = 6\,t^{2}  \Delta^{(\pm)}_{h}( t) + 6\,t^{2} (m_{\eta}^{2} \pm 4\,m_{\eta} m_{\pi} - m_{\pi}^{2} + t)  \Delta_{a}( t)  \nonumber \\
  &\quad\quad \quad  +  \frac{ 3\,t^{2} ( m_{\eta}^{2} \pm 4\,m_{\eta} m_{\pi} - m_{\pi}^{2}+t  )^{2}  -U^{2}(t)}{2}  \Delta_{d}^{(\pm)} ( t) ,\nonumber \\
&  \Delta_{h}^{(\pm)} ( t) = \int_{s_{-}(t)}^{s_{+}(t)}  \!\!\!\!\!\!\!\!\ \!\!\!\!\!\!  (C') \     \frac{d s' }{  U(s')  }( s' -(m_{\eta} \pm m_{\pi})^{2} ) \nonumber \\
&\quad\quad \quad  = U(t) \mp 2\,m_{\pi} m_{\eta}  \Delta_{a} ( t).\nonumber
\end{align}

\item $\bf{(L,L')=(2, 2)}$:

\begin{align}
& \Delta_{2,2}(s,t) = \frac{ 3\,t^{2} (2\,s+t-m^{2}_{\eta} -3\,m^{2}_{\pi} )^{2}  -U^{2}(t)}{2}\Delta_{2,0}(s,t)   \nonumber \\
& \quad\quad\quad\quad   + 6\,t^{2}  \Delta_{n} (s,t)  + 6\,t^{2} (2 s + t - m_{\eta}^{2} -3\,m_{\pi}^{2}) \Delta_{o} (t) ,\nonumber\\
&\Delta_{n}  (s,t) = \int_{s_{-}(t)}^{s_{+}(t)}  \!\!\!\!\!\!\!\!\ \!\!\!\!\!\!  (C') \ d s'  \frac{1}{s'}(1-\frac{s}{s'})  \frac{1 }{  U^{3}(s')  }     \nonumber \\
 & \quad \quad \quad\quad \times  \frac{ 3\,{s'}^{2} (2\,t+s'-m^{2}_{\eta} -3\,m^{2}_{\pi} )^{2}  -U^{2}(s')}{2}  \nonumber \\
   &\quad\quad  \quad  =  \frac{3}{2} \left [  \Delta_{d}^{(-)} (t) - (s-  (m_{\eta} - m_{\pi})^{2})\Delta_{k}^{(-)} (t)  \right ]   \nonumber \\
   & \quad\quad \quad   + 6\,(t + m_{\pi}(m_{\eta} - m_{\pi})) \nonumber \\
   & \quad \quad\quad   \times  \left [ \Delta_{k}^{(-)} (t)  - (s-  (m_{\eta} + m_{\pi})^{2}) \frac{  \Delta_{e}  (t)   -  \Delta_{k}^{(-)} (t) }{4\,m_{\eta}\, m_{\pi}} \right ] \nonumber \\
   &\quad\quad \quad  + 6\,(t + m_{\pi}(m_{\eta} - m_{\pi}))^{2} \bigg [   \frac{ \Delta_{k}^{(+)} (t)  - \Delta_{e}  (t)  }{4\,m_{\eta}\,m_{\pi}}      \nonumber \\
   & \quad \quad\quad \quad  + (s-  (m_{\eta} - m_{\pi})^{2}) \frac{ 2\Delta_{e}  (t) -  \Delta_{k}^{(+)} (t) - \Delta_{k}^{(-)} (t)  }{(4\,m_{\eta}\, m_{\pi})^{2}} \bigg ] \nonumber \\
    &\quad\quad  \quad  -\frac{1}{2} \left [  \Delta_{d}^{(+)} (t) - (s-  (m_{\eta} - m_{\pi})^{2})   \Delta_{e}  (t)    \right ] \nonumber \\
    & \quad\quad  \quad + 2\,m_{\pi}^{2}   \left [ \Delta_{e}  (t)    -  s   \Delta_{c} (0,t)\right ],
\nonumber
\end{align}
\begin{align}
\Delta_{o} (s,t)& = \int_{s_{-}(t)}^{s_{+}(t)}  \!\!\!\!\!\!\!\!\ \!\!\!\!\!\!  (C') \ d s'  \frac{1}{{s'}^{2}}   \frac{1 }{  U^{3}(s')  }    \nonumber \\
&   \times   \frac{ 3\,{s'}^{2} (2\,t+s'-m^{2}_{\eta} -3\,m^{2}_{\pi} )^{2}  -U^{2}(s')}{2}  \nonumber \\
  &   =  \frac{3}{2}   \Delta_{k}^{(-)} (t)    + 6\,(t + m_{\pi}(m_{\eta} - m_{\pi}))   \frac{ \Delta_{e}  (t)  -  \Delta_{k}^{(-)} (t) }{4\,m_{\eta}\,m_{\pi}}  \nonumber \\
  &   - 6\,(t + m_{\pi}(m_{\eta} - m_{\pi}))^{2}    \frac{ 2 \Delta_{e}  (t)  -  \Delta_{k}^{(+)} (t) - \Delta_{k}^{(-)} (t)  }{(4\,m_{\eta}\,m_{\pi})^{2}}  \nonumber \\
  &   - \frac{1}{2}  \left ( \Delta_{e}  (t)  -4\,m_{\pi}^{2}\, \Delta_{c} (0,t)\right ).\nonumber
\end{align}

\end{itemize}

\bibliographystyle{prsty}
\bibliography{Eta_paper_2}

\end{document}